\documentclass{aastex}
\newif\ifgrayscale

\grayscalefalse 

\newcommand{\tnm}[1]{{\tablenotemark{#1}}}
\newcommand{\tnt}[2]{{\tablenotetext{#1}{#2}}}

\newcommand{\eg}{{\it e.g.}}
\newcommand{\ie}{{\it i.e.}}
\newcommand{\etc}{{\it etc.}}
\newcommand{\etal}{{\it et~al.}}

\newcommand{\half}{{1\over2}}

\newcommand{\dif}{\mathrm{d}}

\newcommand{\aei}{(a,e,i)}

\newcommand{\sn}{$S/N$}

\newcommand{\AU}{\,\mathrm{AU}} 
\newcommand{\au}{\,\mathrm{au}} 
\newcommand{\km}{\,\mathrm{km}}
\newcommand{\kms}{\,\mathrm{km}/\mathrm{s}}

\newcommand{\meter}{\,\mathrm{m}}
\newcommand{\cm}{\,\mathrm{cm}}
\newcommand{\mm}{\,\mathrm{mm}}
\newcommand{\um}{\,\mu \mathrm{m}}

\newcommand{\days}{\,\mathrm{d}}
\newcommand{\dayperyear}{\,\mathrm{d}/\mathrm{yr}}

\newcommand{\hour}{\,\mathrm{h}}
\newcommand{\hourperday}{\,\mathrm{h}/\mathrm{d}}
\newcommand{\minute}{\,\mathrm{min}}
\newcommand{\second}{\,\mathrm{s}}

\newcommand{\mags}{\,\mathrm{mag}}
\newcommand{\K}{\,\mathrm{K}}

\newcommand{\kg}{\,\mathrm{kg}}
\newcommand{\g}{\,\mathrm{g}}

\newcommand{\TC}{{2008~TC$_3$}}
\newcommand{\RH}{{2006~RH$_{120}$}}
\newcommand{\XB}{{2012~XB$_{112}$}}

\newcommand{\wps}{\ensuremath{w_{\rm P1}}}

\newcommand{\PSone}{\protect \hbox {Pan-STARRS1}}
\newcommand{\PStwo}{\protect \hbox {Pan-STARRS2}}

\begin{document}

\title{Detecting Earth's Temporarily-Captured
Natural Satellites\\ --- Minimoons}

\author{
Bryce Bolin\altaffilmark{1,2} (bolin@ifa.hawaii.edu),
Robert Jedicke\altaffilmark{1},
Mikael Granvik\altaffilmark{3},
Peter Brown\altaffilmark{4},
Ellen Howell\altaffilmark{5},
Michael C. Nolan\altaffilmark{5},
Peter Jenniskens\altaffilmark{6},
Monique Chyba\altaffilmark{7},
Geoff Patterson\altaffilmark{7},
Richard Wainscoat\altaffilmark{1}
}

\received{December 22, 2013}
\accepted{May 20, 2014}

\slugcomment{44 Pages, 14 Figures, 1 Table}

\altaffiltext{1}{University of Hawaii, Institute for Astronomy, 2680 Woodlawn Dr, Honolulu,
  HI, 96822}
  \altaffiltext{2}{University of Phoenix, Hawaii Campus, 745 Fort St, Honolulu, HI 96813}
\altaffiltext{3}{Department of Physics, P.O. BOX 64, 00014 University of Helsinki, Finland}
\altaffiltext{4}{University of Western Ontario, Physics \& Astronomy Department, London, Ontario, Canada}
\altaffiltext{5}{Arecibo Observatory, Arecibo, Puerto Rico}
\altaffiltext{6}{SETI Institute, Carl Sagan Center, 189 Bernardo Avenue, Mountain View, CA 94043}
\altaffiltext{7}{University of Hawaii, Department of Mathematics, Honolulu,
  HI, 96822}

\shorttitle{Detecting Earth's Minimoons}
\shortauthors{Bolin \etal}


\begin{abstract}
We present a study on the discoverability of temporarily
captured orbiters (TCOs) by present day or near-term anticipated ground-based and space-based
facilities. TCOs \citep{Granvik2012} are potential targets for spacecraft
rendezvous or human exploration \citep{Chyba2014} and provide an
opportunity to study the population of the smallest asteroids in
the solar system. We find that present day ground-based optical
surveys such as Pan-STARRS and ATLAS can discover the largest TCOs
over years of operation.  A targeted survey conducted with the Subaru
telescope can discover TCOs in the $0.5\meter$ to $1.0\meter$ diameter
size range in about 5 nights of observing. Furthermore, we discuss the
application of space-based infrared surveys, such as NEOWISE, and
ground-based meteor detection systems such as CAMS, CAMO and ASGARD in
discovering TCOs.  These systems can detect TCOs but at a
uninteresting rate.  Finally, we discuss the application of bi-static
radar at Arecibo and Green Bank to discover TCOs.  Our radar
simulations are strongly dependent on the rotation rate distribution
of the smallest asteroids but with an optimistic distribution we find
that these systems have $>80$\% chance of detecting a $>10\cm$
diameter TCO in about 40$\hour$ of operation.
\end{abstract}


\maketitle


\clearpage

\mbox{ }

\vspace{0cm}

\noindent {\bf Proposed Running Head:} Detecting Earth's Natural Satellites \\

\vspace{10cm}

\noindent {\bf Editorial correspondence to:} \\
Bryce Bolin \\
Institute for Astronomy \\
University of Hawaii \\
2680 Woodlawn Drive \\
Honolulu, HI 96822 \\
Phone: +1 808 294 6299 \\
Fax: +1 808 988 2790 \\
E-mail: bolin@ifa.hawaii.edu


\section{Introduction}
\label{s.Introduction}

\citet{Granvik2012} suggested that there exists a steady
state population of natural irregular satellites of the Earth.  We will call these objects temporarily captured orbiters (TCO) or minimoons.  There is almost no previous work on TCOs and what little exists is
reviewed in \citet{Granvik2012}.   They are captured from a
dynamically suitable subset of the near-Earth object (NEO) population
--- those that are on Earth-like orbits with semi-major axis
$a\sim1.0\AU$, eccentricity $e\sim0.0$ and inclinations $i\sim0\deg$
--- and complete an average of $\sim2.9$ revolutions around
Earth during their average capture duration of about 290\,days or 9.5\,months.
In this work we evaluate options for detecting the TCOs as they are
being captured, while they are on their geocentric trajectories, and
in their meteor phase (about 1\% of TCOs enter Earth's atmosphere).

The first confirmed TCO, \RH, was
discovered in 2006 by the Catalina Sky Survey \citep{Larson1998,
  Kwiatkowski2009}.  Its pre- and post-capture orbit, geocentric
trajectory, size, and TCO lifetime were consistent with the
\citet{Granvik2012} model (that did not include information about \RH).
\RH\ had an absolute magnitude of $H=29.9\pm0.3$ corresponding to a
diameter in the range of 2 to 6~meters if we assume S- and C-class
albedos ($p_v$) of 0.26 and 0.064 respectively \citep{Mainzer2012}.  This size range is consistent with the preliminary
estimate of $>2.3\meter$ diameter from the Goldstone radar
facility\footnote{L.~Benner, personal communication.}.  For comparison, \citet{Granvik2012} suggest
that the largest TCO in the steady state population is in the 1 to 2\,m
diameter range, that objects in the 5 to 10$\meter$ diameter range are captured every
decade, and a $100\meter$ diameter TCO is captured about every
100,000\,years.  Thus, the predicted flux of objects in \RH's size
range is in rough agreement with the length of time that optical
telescopic surveys have been in regular operation and their
time-averaged sensitivity to objects like \RH.

However, there is no {\it a priori} reason to expect that the actual
TCO orbit and size-frequency distribution (SFD) should match the
\citet{Granvik2012} model because they used the \citet{Bottke2002}NEO orbit model that is strictly
applicable only to much larger objects of $\ga100\meter$ diameter and
 made no allowance for
non-gravitational forces acting on the small NEOs 
\citep[\eg\ the Yarkovsky effect][]{Morbidelli2003}.  \citet{Granvik2012}'s favored TCO SFD from
\citet{Brown2002} for bolides is appropriate for TCOs in the $0.1\cm$
to $1\meter$ diameter range but the TCO orbit distribution should be
considered suspect without accounting for the radiation forces.

Furthermore, the orbit distribution from the \citet{Bottke2002} NEO
model has several known problems including but not limited to 1)
underestimating the fraction of the NEO population with perihelion
$q<1.0$\,AU, 2) underestimating the fraction of objects on low
inclination orbits \citep{Greenstreet2012} and 3) having coarse
resolution in \aei-space.  These issues are particularly
important because they make it difficult to estimate the population of
dynamically capturable NEOs.  For instance, the entire set of
pre-capture TCO orbits is contained within just 4 bins in the
\citet{Bottke2002} NEO model.  Thus, measuring the TCO size and orbit distribution provides a
sensitive means of testing current NEO and meteoroid SFD models in a
size range that transitions between the two regimes.  

The problem with discovering and characterizing the smallest NEOs is
that they are discovered close to Earth when they are moving so fast
that there is almost no time to coordinate followup during the time
when they are brightest.  The smallest known NEO with $H\sim33.2$
\citep[2008~TS$_{26}$;][]{mpec2008ts26} corresponds to a sub-meter
diameter meteoroid\footnote{To facilitate the conversion between
  absolute magnitude and diameter we use an albedo of $p_v\sim 0.11$
  throughout this work. This albedo results in a 1\,km diameter object
  with $H=18$, a 1\,m diameter has $H=33$, and a 10\,cm diameter
  object has $H=38$.  Our choice makes a difference of only about 13\%
  in the calculated diameter at any absolute magnitude relative to the
  `usual' assumption of $p_v=0.14$ --- well below the systematic error
  in the NEO and TCO models and the systematic errors intrinsic to our
  TCO SFD assumption}.  The smallest NEO with a measured period is
\RH, and the smallest NEO with measured colors/spectrum is 2008~TC$_3$
\citep{Jenniskens2009}, the first object that was discovered prior to
impact.  TCOs offer an opportunity to obtain good physical
characterization of small asteroids because some will be visible for
days or weeks and their location can be predicted accurately.
Measuring their spin rates over a long temporal baseline may enable a
determination of the YORP effect \citep[\eg][]{Bottke2006} and/or the
detection of the change in an asteroids's spin properties during close
and slow Earth encounters \citep[\eg][]{Moskovitz2013}.  Additionally,
their taxonomic classification will have implications for the relative
delivery rates of objects in this size range from the main belt
\citep[\eg][]{Bottke2002a}.

Pre-impact observations of impactors has value extending beyond
improving asteroid-meteorite linkages as exemplified by the case of
\TC \citep{Jenniskens2009} and the Almahatta Sitte meteorite
\citep{Jenniskens2010}. More generally, since all observations of
meteoroids impacting Earth use the atmosphere as a detector, there is
an urgent need for calibration, particularly of mass, across all
meteor measurement techniques. This is a challenging problem as the
physical nature of the meteoroid prior to impact is almost always
unknown (\TC\ being the one exception). Attempts to calibrate the
optical meteoroid mass scale in the 1960s using gram-sized artificial
meteors \citep{Ayers1970} were limited to disk or cone-shaped pure
iron-nickel materials for operational simplicity, which limited their
applicability to the question of pre-atmospheric masses of highly
porous, randomly shaped, non-iron meteoroids.  No similar artificial
meteor measurements have been performed for any other measurement
technique (\eg\ radar meteor calibration). Natural meteorites have
been used to calibrate the mass of their parent meteoroids for $\sim
20$ fireballs \citep[\eg][]{Brown2011,Popova2011} but the technique
\citep{Ceplecha2005} is only applicable to meteoroids with masses of
tens of kilograms.

TCOs could provide the calibration standards if enough of them can be
discovered because 1\% of them eventually become low-speed meteors
\citep{Granvik2012}. As the impact location of a TCO would be known in
advance, ground deployment of suitable sensors could be undertaken
allowing for meteoroid-meteor calibration campaigns.  Moreover, knowledge of the shape and rotation state of a TCO prior to
impact could help resolve the ongoing debate over whether time-varying
optical meteor emission \citep[\eg][]{Beech2000} and received radar
power \citep{Kero2005} is due to meteoroid rotation modulating
ablation or some other mechanism \citep{Babadzhanov2004}.

Finally, TCOs are likely the lowest $\Delta v$ natural targets
for spacecraft missions \citep{Granvik2013b,Granvik2012,Elvis2011}.  The
opportunity to retrieve kilograms or even hundreds of kilograms of
pristine asteroidal material unaffected by passage through
Earth's atmosphere and uncontaminated by exposure to elements on
Earth's surface would be a tremendous boon to solar system studies.
We can even imagine multiple retrieval missions to TCOs that have been
taxonomically pre-classified as interesting targets --- essentially
sampling the range of the main belt in our own backyard.


\section{Observable characteristics of the steady state TCO population}
\label{s.SteadyStateTCOPopulation}

\citet{Granvik2012} provide trajectories (orbits) for about
18,000~TCOs that are designed to be a representative, unbiased,
population.  We have used this synthetic TCO population to calculate the observable properties of the steady state TCO population as described in detail in the appendix.

\subsubsection*{Geocentric TCO spatial distribution}

\begin{figure}
\centering
\hspace*{-3.72cm}
\ifgrayscale
\includegraphics[scale=0.43]{TCO_r_theta_V3_BW-eps-converted-to.pdf}
\else
\includegraphics[scale=0.43]{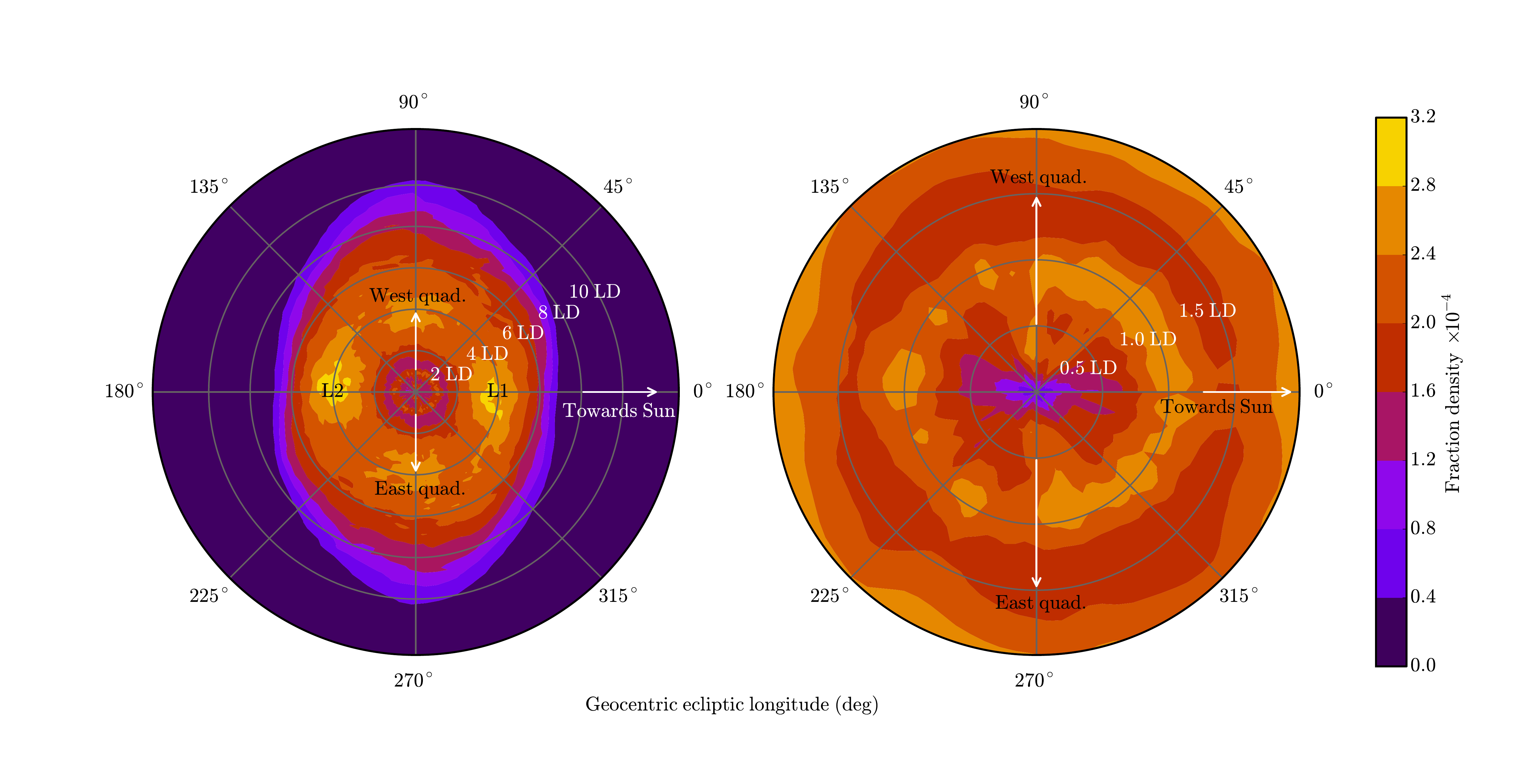}
\fi
\caption{Normalized geocentric TCO
  $\Delta-\theta$ residence density distribution
  (eq.~\ref{eq.normalized_residence}) with no restrictions on apparent
  magnitude or rate of motion where $\Delta$ is the geocentric
  distance and $\theta$ is the azimuthal angle about the ecliptic pole
  with the origin in the sunward direction.  The panel on the right is
  a magnified version of the panel on the left for $\Delta<2$\,LD. The
  Sun is to the right in both panels \ie\ at $0\arcdeg$.  The number
  of objects in the $\Delta$-$\theta$ range,
  $(\Delta\pm\dif\Delta/2,\theta\pm\dif\theta/2)$, is the product of
  the fraction density and $(\dif\Delta,\dif\theta)$.}
\label{fig.TCO_rtheta_residency_noconstraints}
\end{figure}

Fig.~\ref{fig.TCO_rtheta_residency_noconstraints} is useful for
understanding the observable characteristics of TCOs with special
relevance to their detection by radar or spacecraft.  Note that there are high TCO density patches near the
Sun-Earth L1 and L2 points and also at 4\,LD (lunar distance,
$\sim0.00257\au$) in the direction towards east and west
quadrature.  The largest steady-state TCOs are just barely visible at
opposition at this distance even with large aperture ground-based visible light
surveys.  Their overall spatial distribution from this viewpoint is
oval with the elongation perpendicular to the Sun-Earth direction.
Some of the minimoons can extend to just beyond 10\,LD in the
direction of both quadratures.  The magnified
distribution in the right panel shows an enhancement of objects at the
Moon's distance with a `moat' in the density at $\sim1.5$\,LD.  There
is also an enhancement of TCOs near Earth at the quadratures --- while
the TCOs may be close to Earth in this orientation their phase angle
will be $\sim90\arcdeg$ and they will appear much fainter in visible
light than at the same distance in the direction towards opposition.
Unfortunately, the TCOs are depleted in the anti-Sun direction out to
and beyond the Moon's distance.

\begin{figure}
\centering
\hspace*{-0.302cm}
\ifgrayscale
\includegraphics[scale=0.6]{TCO_geocentric_distance_v2-eps-converted-to.pdf}
\else
\includegraphics[scale=0.6]{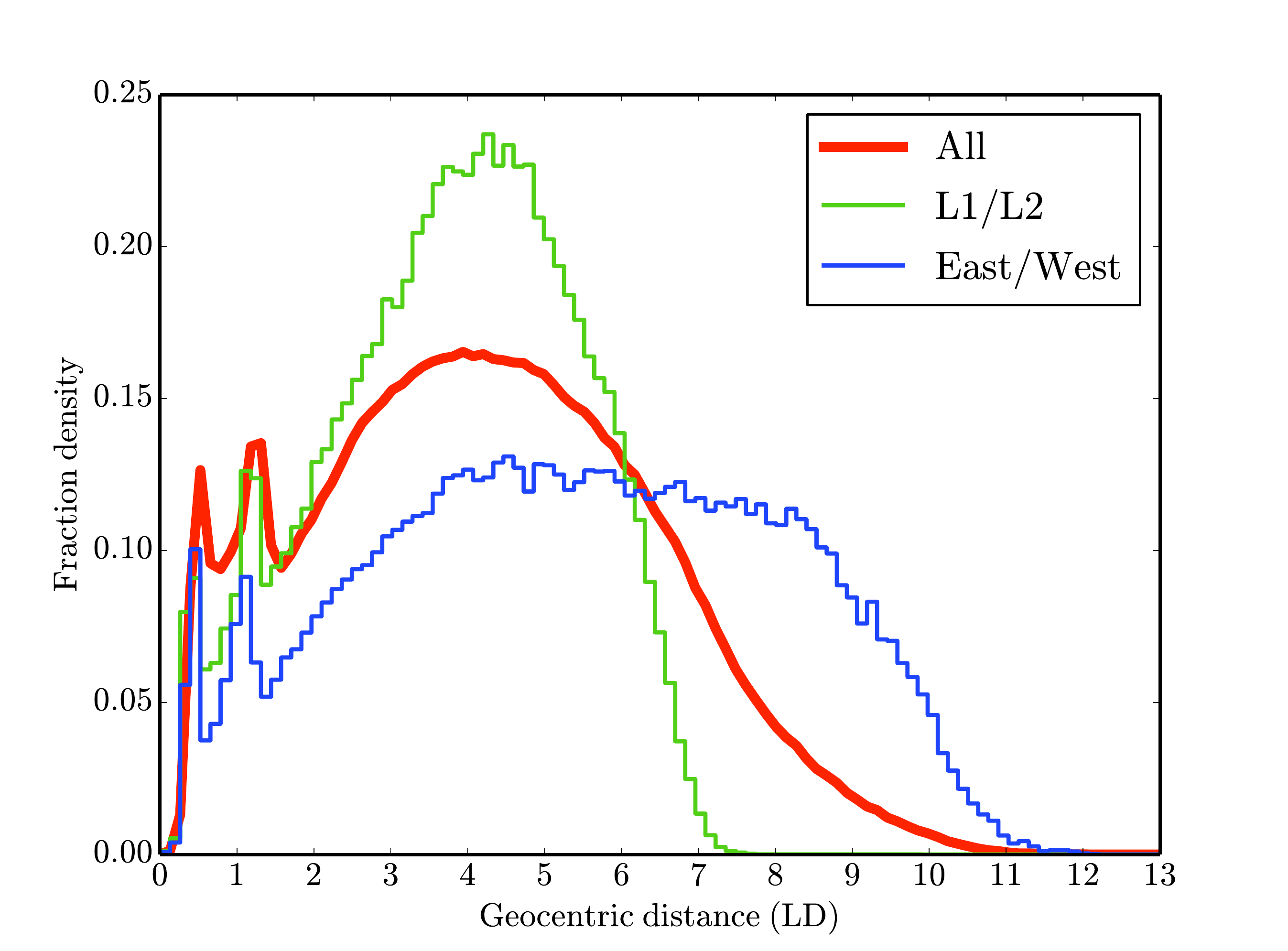}
\fi
\caption{Geocentric distance distribution in lunar distances
  ($\sim0.00257\au$) for all TCOs (thick black line), near the L2 and L1 Lagrange points (light gray line),
  and near the two quadratures (dark gray line).  `Near' means within 15$\arcdeg$ in
  longitude and within 10$\arcdeg$ in latitude. The number of objects
  in the geocentric distance range $(\Delta\pm\dif\Delta/2)$ is the
  product of the fraction density and $\dif\Delta$.}
\label{fig.TCO_geocentric_distance}
\end{figure}

The TCO geocentric distance distribution illustrated in
fig.~\ref{fig.TCO_geocentric_distance} is critical to assessing
opportunities for detecting and even discovering TCOs at radar
facilities (see \S\ref{s.radarDetection}).  There is no significant
difference between the distributions in the L1 and L2 directions or in
the east and west quadratures but there is a clear difference between
the two sets.  The average geocentric distance is $4.3\pm2.2$(rms)~LD
for the L1/L2 regions, $5.7\pm2.9$(rms)~LD for the regions near the
east and west quadratures, and $4.1\pm1.6$(rms)~LD for the entire TCO
population.  The differences work in favor of optical detection of the
TCOs since they are closer to Earth near opposition but a 1\,m
diameter object with absolute magnitude $H \sim 33$ will have an
apparent magnitude of $V=23.1$ at the average opposition distance. The
peaks at $\sim0.5$ and $\sim1.5$\,LD are caused by long-lived particles
that have lifetimes of $>2,000\days$.

\subsubsection*{Geocentric TCO sky plane distribution}

\begin{figure}
\centering
\hspace*{-0.302cm}
\ifgrayscale
\includegraphics[scale=0.5]{TCO_skyplane_residency_noconstraints_v3-eps-converted-to.pdf}
\else
\includegraphics[scale=0.5]{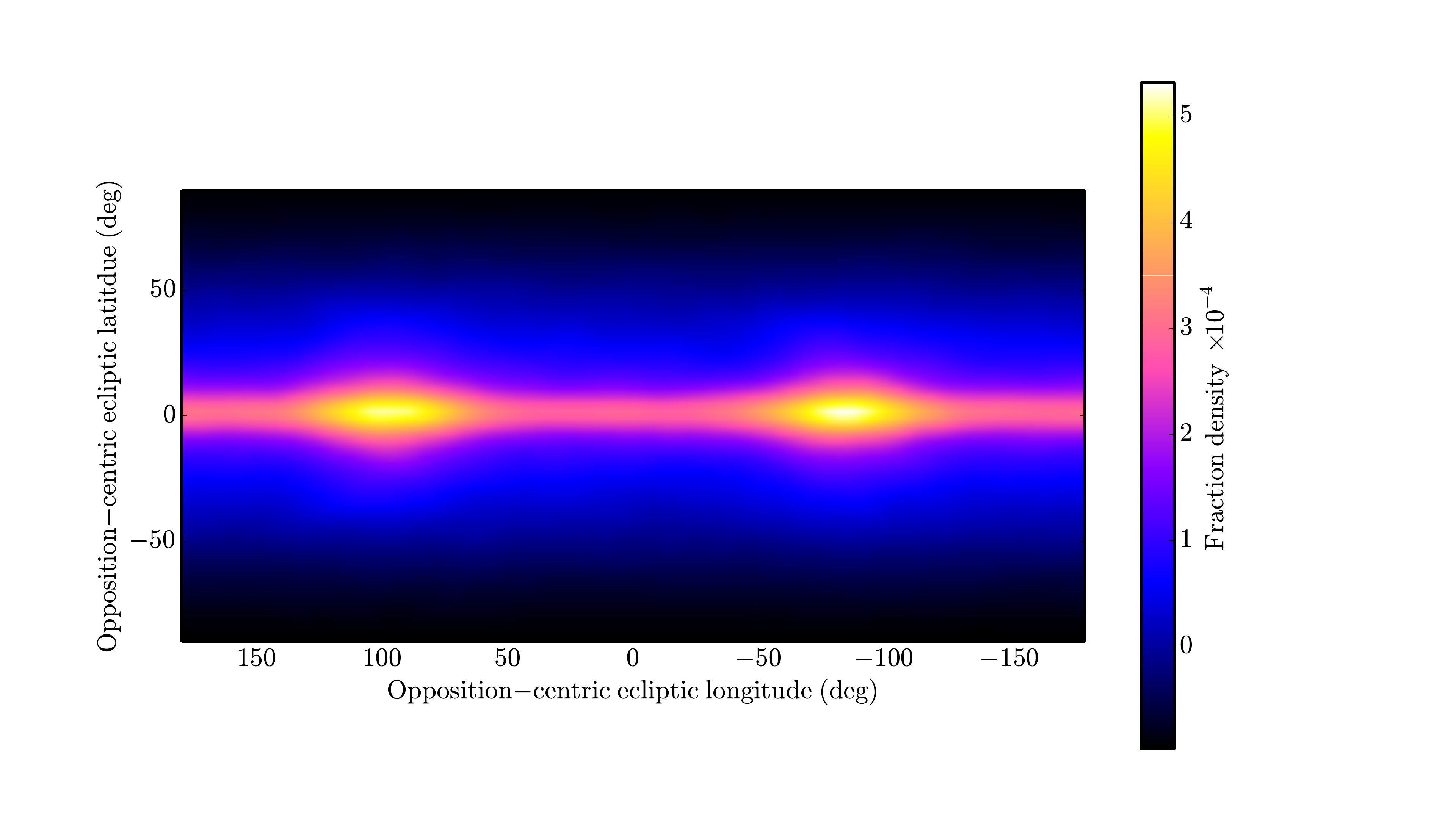}
\fi
\caption{Normalized geocentric TCO sky plane residence density
  distribution (eq.~\ref{eq.normalized_residence}) with no
  restrictions on apparent magnitude or rate of motion in an
  opposition-centric ecliptic reference system.  Negative
  opposition-centric longitudes are west of opposition.  The values
  are the fraction of the population in $3\arcdeg \times 3\arcdeg$
  bins. The number of objects in the opposition-centric longitude and
  latitude range, $(\lambda\pm\dif\lambda/2,\beta\pm\dif\beta/2)$, is
  the product of the fraction density and $(\dif\lambda,\dif\beta)$.}
\label{fig.TCO_skyplane_residency_noconstraints}
\end{figure}

\citet{Granvik2012} provide trajectories (orbits) for about
18,000~TCOs that are designed to be a representative, unbiased,
population.  \ie\ they represent the steady state distribution of TCOs
around Earth.  Their sky plane residence time distribution (defined in \S\ref{appendix.SkyPlaneResidenceTimeDistribution}) without any
constraints on their apparent brightness, distance, rate of motion,
\etc, shows a strong enhancement near east and west quadrature (see
fig.~\ref{fig.TCO_skyplane_residency_noconstraints}.)  Their
trajectories\footnote{We often refer to TCO motion around Earth as a
  trajectory because their motion often does not follow a
  nearly-closed elliptical path like the conventional notion of an
  `orbit'.} are such that they tend to be furthest from Earth, and
therefore moving slowly, at both quadratures so they
spend more time there and have the highest sky plane density at that
location.  The TCOs are strongly enhanced on the ecliptic with the sky
plane density dropping off rapidly with ecliptic latitude as for most
classes of objects in the solar system as viewed from Earth.

\begin{figure}
\centering
\hspace*{-0.142cm}
\ifgrayscale
\includegraphics[scale = 0.55]{TCO_skyplane_residency_LSST_v2-eps-converted-to.pdf}
\else
\includegraphics[scale = 0.55]{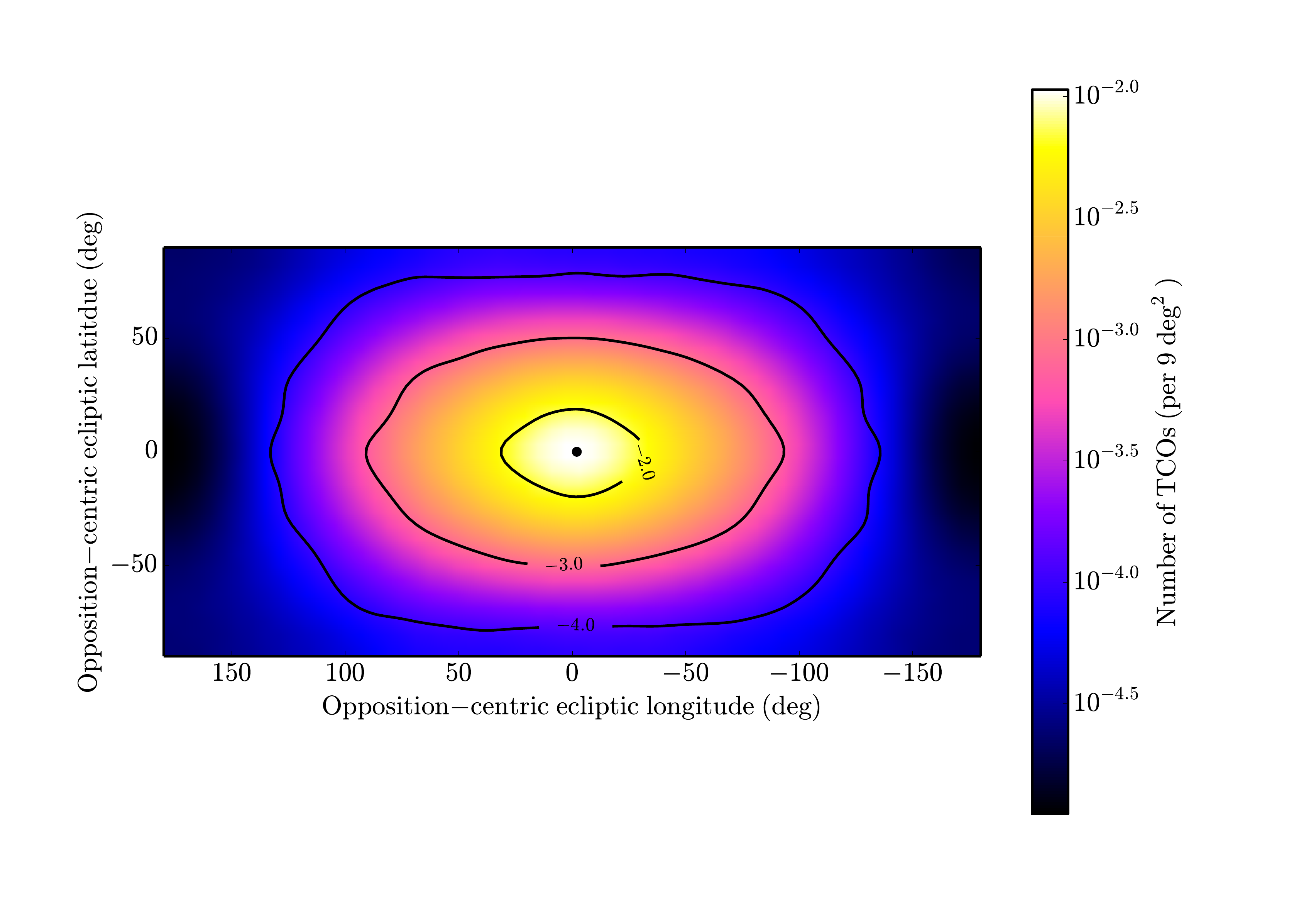}
\fi
\caption{Sky plane number density of TCOs with $H<38$ ($\sim10\cm$
  diameter), apparent magnitude $V<20$ and rate of motion
  $<15\arcdeg/\days$.  The constraints are roughly consistent with the
  expected performance characteristics of the ATLAS system
  \citep{Tonry2011} but are generally applicable to any telescopic TCO
  survey except for (roughly) a normalization constant.  The small
  black circle in the center represents the size of Earth's umbra and
  penumbra at 4 lunar distances.  The values are the number of TCOs in
  $3\arcdeg \times 3\arcdeg$ bins.}
\label{fig.TCO_Normalized_Skyplane_distribution_rate_of_motion_V_lt_24.7_rate_lt_10}
\end{figure}

The sky plane number density with constraints on the TCOs's magnitude
and rate of motion shown in
fig.~\ref{fig.TCO_Normalized_Skyplane_distribution_rate_of_motion_V_lt_24.7_rate_lt_10}
is dramatically different from the unconstrained sky plane
distribution.  Application of the IAU standard $H$ and $G_{12}$
asteroid photometric model \citep[][we use $G_{12}=0.5$ throughout
  this work]{Muinonen2010} eliminates the enhancements at the east and
west quadratures observed in
fig.~\ref{fig.TCO_skyplane_residency_noconstraints} in favor of a
strong enhancement at opposition due to the photometric surge for very
small phase angles.  The enhancement along the ecliptic is no longer
present because the TCOs must be close to Earth to be detectable which
has the effect of increasing their apparent ecliptic latitude
distribution.  Thus, it is clear that a telescopic optical TCO survey
is most effective towards opposition but need not concentrate on the
region near the ecliptic.  Near-opposition surveying is typically
employed by contemporary NEO surveys
\citep[\eg][]{Jedicke1996,Jedicke2002} so we expect that the NEO
survey strategy should be finding TCOs or that the next generation of
sky surveys will do so as a matter of course
\citep[\eg][]{Tokunaga2007}.

All our results that require calculating the TCOs' apparent brightness
account for Earth's umbral and penumbral shadowing.  Earth's umbral
shadow is about $1\arcdeg$ in diameter at 1\,LD and is always in the
direction of opposition, exactly where the TCO sky plane density is
highest.  We assume that within the umbral shadow there is zero
sunlight and we reduce a TCO's
apparent brightness by $2.5\log f_\Sun\mags$ in the penumbra where $f_\Sun$ is
the fraction of the Sun's `surface' visible at the TCO.  We found that
Earth's shadow makes only a 4\% difference in the total sky plane
residence time for the detection parameters in
fig.\,\ref{fig.TCO_Normalized_Skyplane_distribution_rate_of_motion_V_lt_24.7_rate_lt_10}
and a 5\% decrease in the $3\arcdeg\times3\arcdeg$ bin at opposition.
The modest shadowing effect even for the relatively bright limiting
magnitude is a consequence of requiring that the rate of motion be
$<1\arcdeg/\days$.  The Moon's rate of motion is about
$12\arcdeg/\days$ so most of the TCOs in the figure must be near or
beyond the Moon and mostly unaffected by Earth's shadow.  If an
optical survey was capable of detecting faster rates of motion then
the Earth shadowing would have a larger effect on the TCO detection
capability.

\subsubsection*{Geocentric TCO apparent rates of motion}

In designing a TCO survey strategy (see \S\ref{s.OpticalDetection}) it
is important to understand the TCOs' rates of motion.  Objects that
move by more than about 2 PSF-widths during the course of an exposure
spread their total flux in a `trail' that reduces the peak \sn\ in any
pixel along the trail compared to stationary objects of the same
intrinsic brightness \citep[\eg\ ][]{Veres2012}.  Peak pixel detection algorithms are therefore
less likely to identify the trailed source.  Even algorithms that can
identify trailed detections, \eg\ by summing flux along a line in the
image, have reduced sensitivity because the trail's total \sn\ is also
less than the overall \sn\ for a stationary object of the same
intrinsic brightness.  In this work we partly ignore trailing effects
because they are detection-system and software-algorithm dependent.
We simply assume that the system will be able to detect objects that
are not moving too fast as long as their apparent magnitude is
brighter than the system's limiting value.\footnote{For more
  information on trail fitting see \eg\ \citet{Veres2012} who provide
  a technique for fitting trailed source detections but not how to
  identify them.}

TCOs within about $30\arcdeg$ of opposition with $V<20$ and rates of
motion $<15\arcdeg/\days$ have an average apparent angular speed of
about $10\arcdeg/\days$.  (For comparison, the only currently known
minimoon, \RH, was discovered about $5\arcdeg$
from opposition moving at $\sim 10\arcdeg / \days$ with $V\sim
19\mags$).  Typical exposure times for contemporary sky surveys are on
the order of 60\,s so the average TCO would move about $25\arcsec$
during the exposure.  This could hamper TCO detection by survey
systems with small PSFs and pixels because the TCO flux will be
distributed over many pixels and create a long trail.  Conversely,
survey systems with larger PSFs and pixels will be less affected by
the TCOs' motion.  Automated trail-identification software or human
vetting of TCO candidates could alleviate the loss of detections due
to PSF trailing provided that there are not too many image artifacts
that mimic trailed detections.  Thus, large aperture systems with short
exposure and fast readout times would be best for TCO discovery
because the large aperture allows the detection of faint objects while
the short exposures minimizes trailing losses.

\subsubsection*{TCO detection at capture in Earth-Moon system}

\begin{figure}
\centering
\hspace*{-1.0cm}
\ifgrayscale
\includegraphics[scale=0.47]{TCO_at_L2_capture_v3_bb_bw-eps-converted-to.pdf}
\else
\includegraphics[scale=0.7]{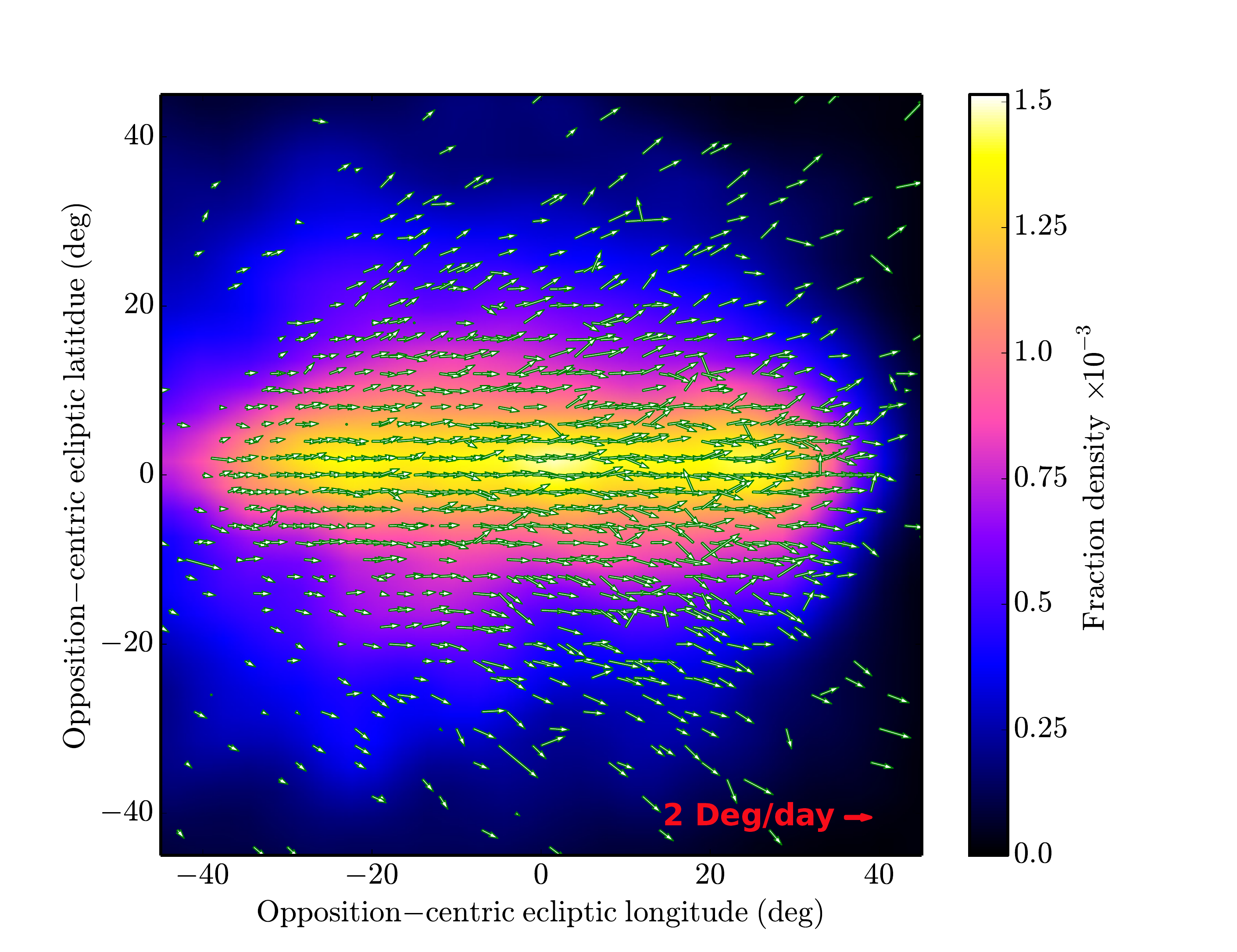}
\fi
\caption{Sky plane distribution of TCOs at the moment of capture near
  L2 without constraints on the apparent magnitude or rate of motion.
  The contour values are the fraction of the population at capture in
  the $3\arcdeg \times 3\arcdeg$ bins.  The vectors represent the
  average rate of motion and direction of the TCOs in the bins at the
  time of capture.  A vector representing a $2\arcdeg/\days$ rate of
  motion is shown in the lower right. The number of objects in the
  opposition-centric longitude and latitude range,
  $(\lambda\pm\dif\lambda/2,\beta\pm\dif\beta/2)$, is the product of
  the fraction density and $(\dif\lambda,\dif\beta)$.}
\label{fig.TCO_at_L2_capture}
\end{figure}

Figure~12 of \citet{Granvik2012} provides an idea for how to design a
targeted TCO survey (see \S\ref{ss.OpticalTelescopeDetection}).  The
figure shows that at the moment of capture, the time at which their
total energy becomes negative with respect to the Earth-Moon
barycenter, most TCOs are 1) in retrograde geocentric orbits 2) near
L1 or L2 and just outside Earth's Hill sphere 3) moving in roughly the
same direction 4) at similar rates of motion.
Figure~\ref{fig.TCO_at_L2_capture} shows that at the time of capture
the TCOs are concentrated near the ecliptic in a $\sim20\arcdeg$ wide
band centered at opposition (the L2 direction).  Indeed, there are
$\sim6$ TCOs larger than 10\,cm diameter captured every day in both L1
and L2 (half at each).  The objects' motion vectors bring them near
the opposition point before or after their moment of capture so that
there is a steady stream of small objects in roughly the same
direction moving at roughly the same rate of motion.  The rates of
motion are relatively modest by NEO standards at
$1.8\pm0.7$(rms)$\arcdeg/\days$ with the objects' motion at an average
position angle of $274\arcdeg\pm38\arcdeg$(rms) \eg\ typically
retrograde consistent with most motion along the ecliptic.  The
trailing rate corresponds to $\sim0.075\arcsec/\second$ so the average TCO
trails by about 1$\arcsec$ in $\sim13$\,s.  However, if the telescope
is tracked at the mean TCO rate then all the objects with rates within
$\pm1$\,rms of the mean will trail by less than 1$\arcsec$ in about
34\,s.  A large telescope with a wide passband filter can reach
$V\sim24.5$ at a good site with $1\arcsec$ seeing so that trailing
losses can be dramatically reduced for the TCOs in exposures of
$\la60$\,s.  Furthermore, other TCOs that have already been captured
may pass through the same region and contribute more detections to
this type of targeted TCO survey.  Thus, a targeted deep survey
towards the opposition point might detect TCOs near the time of
capture and can take advantage of the relatively smooth flow of
objects through the region.

\subsubsection*{Detection of TCOs in the infrared}

\begin{figure}
\centering
\hspace*{-0.80cm}
\ifgrayscale
\includegraphics[scale = 0.5]{TCO_flux_wise_1_m_4_LD-eps-converted-to.pdf}
\else
\includegraphics[scale = 0.55]{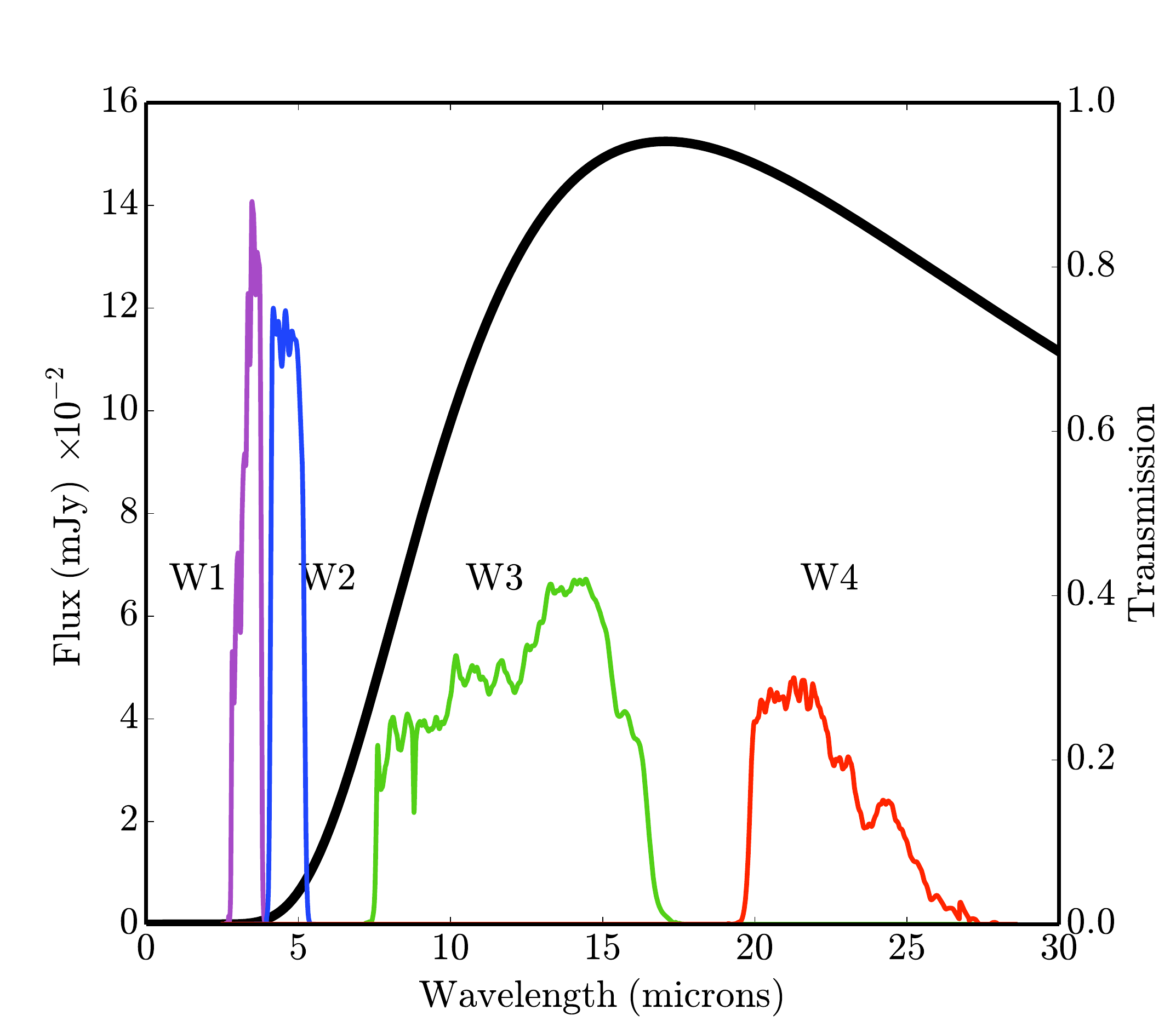}
\fi
\caption{The thick solid line represents the IR flux (left axis) for a
  1 meter diameter minimoon at $300\K$ at 4 LD from the observer.
  (Details of the TCO IR flux calculation are provided in
  \ref{appendix.InfraredDetection}) The thinner curves illustrate the
  transmission percentages (right axis) for the 4 WISE pass bands
  (from {\tt
    http://www.astro.ucla.edu/\textasciitilde{}wright/WISE/passbands.html}).}
\label{fig.TCO_WISE_Flux}
\end{figure}

It may also be possible to detect TCOs in the infrared from
space-based platforms or perhaps detections already exist in the WISE
spacecraft images
\citep[\eg][\S\ref{s.InfraredDetection}]{Mainzer2011b}.  If we assume
that the TCOs in the size range from 10\,cm to 1\,m are small enough
and/or rotating rapidly enough to be in thermal equilibrium then there
is an advantage to detecting the TCOs in the infrared because they are
not affected by phase angle effects.  The peak of the thermal
radiation for a 1\,m diameter blackbody object occurs at about
16\,$\mu$m and the peak integrated signal occurs
in the 12\,$\mu$m W3 passband of the 4 WISE filters (see fig.~\ref{fig.TCO_WISE_Flux}). Thus, the sky plane TCO number
distribution in the IR does not have any strong enhancements along the
ecliptic as illustrated in fig.~\ref{fig.TCO_12micron_skyplane_residency}.
Thus, a near-geocentric IR TCO survey (\ie\ from the ground or
relatively low-Earth orbit) would be best served by surveying along
the ecliptic as much as possible and then expanding the search to
higher latitudes.  However, we think a space-based IR survey would be
best located at the Earth-Sun L1 Lagrange point from which the majority of
TCOs are within a smaller area on the sky in the direction toward
Earth (see fig.~\ref{fig.TCO_12micron_skyplane_residency}).

\subsubsection*{Detection of TCO meteors}

Finally, we consider detecting TCOs in the final moments of their
geocentric trajectory as they enter Earth's atmosphere (see
\S\ref{ss.OpticalMeteorSurveyDetection} and
\S\ref{ss.MeteorRadarDetection}) as \citet{Granvik2012} showed that
about 1\% of all TCOs become meteors.  These TCO meteors strike the
Earth's atmosphere in a narrow range of speeds with an average of
$11.19\pm0.03$(rms)\,km/s --- nearly equal to Earth's escape speed
because the TCOs have essentially fallen to Earth from a large
distance with $v_\infty\sim 0\kms$.  The impacting TCOs have a flat
distribution in $\sin \theta_i$ where $\theta_i$ is the impact angle,
the acute angle between the TCO's trajectory and the perpendicular to
the Earth's surface.  Thus, the mean impact angle for the TCOs in
\citet{Granvik2012}'s sample is $44.6\arcdeg\pm 21.4\arcdeg$(rms).


\section{Optical Detection}
\label{s.OpticalDetection}

In this section we consider different instances of multiple methods of
detecting TCOs in optical light: 1) all-sky, wide area and targeted
telescopic surveys and 2) meteor surveys.  We make the distinction
between all-sky and wide-area because within the next few years the
Asteroid Terrestrial-impact Last Alert System (ATLAS)
\citep{Tonry2011}, and possibly the US Department of Defense's Space
Surveillance Telescope (SST), will survey the entire night sky
multiple times each night.  Other contemporary and anticipated surveys
only cover wide areas of the sky each night.  The meteor surveys
typically either 1) monitor the entire sky visible from their
locations each night but necessarily to a relatively bright limiting
magnitude \citep[\eg][]{Spurny2007} or 2) perform a deeper survey over
narrower fields, often using image intensified video systems
\citep{Hawkes2002,Weryk2013}. A meteoroid orbit survey described in
\citet{Jenniskens2011} accomplishes both goals using a `fly's eye'
approach with low-light video cameras and multi-station imaging.

Optical surveys with limited coverage should target the region near opposition since TCOs are enhanced in the optical in the anti-Sun direction (see fig.~\ref{fig.TCO_Normalized_Skyplane_distribution_rate_of_motion_V_lt_24.7_rate_lt_10}), and expand their TCO search outwards from opposition as the survey permits.  Surveys with fainter limiting magnitudes will have
higher TCO discovery rates and those with software capable of
identifying the trailed TCO detections will also have an advantage
over those surveys that can only identify detections with stellar
PSFs.

\subsection{Telescopic Detection}
\label{ss.OpticalTelescopeDetection}

We considered in detail four different instances of optical telescopic surveys
with which we have some familiarity and focussed on next-generation
surveys with wide field coverage or a targeted survey to a faint
limiting magnitude with a next-generation camera (see
table~\ref{t.TCOSurveyPerformance}).

\subsubsection*{Catalina Sky Survey (CSS)}

The only known TCO, \RH, was discovered during routine survey
operations by the Catalina Sky Survey \citep[CSS,\ ][]{Larson1998,
  Kwiatkowski2009} so that survey's TCO detection capabilities have
already been demonstrated.  The CSS has operated 2 or 3 telescopes in
the northern and southern hemispheres for about a decade and is
currently the leading NEO discovery survey.  Their two northern
hemisphere telescopes (observatory codes G96 and 703) serve
the complementary purposes of a relatively deep and narrow survey along
with a shallow and wide survey.  Each of their sites currently
delivers roughly comparable numbers of NEO discoveries to the
\PSone\ system described in more detail below (almost identical for
G96 and within about 50\% for 703).

\subsubsection*{Panoramic Survey Telescope and Rapid Response System (\PSone)}

The \PSone\ survey has been in operation since the spring of 2010.
The 1.8\,m telescope with an $\sim7$\,deg$^2$ field of view discovers
most of its moving objects in 45\,s exposures obtained with their
wide-band $\wps$ filter.  Since the onset of operations they have
surveyed about 7,700\,deg$^2$/month in modes suitable for TCO
detection to a limiting magnitude of $V\sim21.7$.  \PSone\ has not yet
implemented trail detection and currently has essentially zero
efficiency for identifying objects moving faster than about
$3\arcdeg/\days$ \citep{Denneau2013}.  

Since we have more experience with the \PSone\ system we will use it's
{\it current} performance characteristics to model TCO detections but
the results will also be roughly representative of the
CSS G96 site (roughly similar limiting magnitudes and nightly survey area).  Our choice of system characteristics is motivated by our
desire to ensure that 1) we use realistic NEO survey detection
characteristics of an existing survey and 2) our estimates represent a
lower limit to the TCO detection capabilities of the more capable and
upcoming NEO surveys.

In its current mode of operations \PSone\ should detect about $10^{-2}$ TCOs/lunation so it is unsurprising that it has not
yet reported a TCO discovery.  However, it is likely that they have
already imaged trailed TCOs that were not identified by their image
processing pipeline or linked by the moving object processing system
--- a large great circle residual (GCR\footnote{the astrometric RMS of the
  detections relative to the best-fit great circle}) of the detections
within a tracklet relative to the astrometric uncertainty makes it
difficult to link the detections into tracklets.  Furthermore, even if
a TCO was detected and reported to the MPC it is unlikely that it
could be reacquired by followup observatories because of the
relatively large ephemeris uncertainty, which is due to their high
rates of motion, and the fact that \PSone\ typically requires
8-12$\hour$ to report moving objects.  

We expect that these problems
will be dramatically reduced over the next few years as
\PSone\ and CSS are upgraded and other NEO surveys come online.  The implementation of trail-identification
software and decreasing the time delay between image acquisition and
reporting candidate discoveries to the MPC will increase the likelihood of TCO discoveries.  Starting in April 2014 the \PSone\ system was
dedicated 100\% of the time to NEO surveying and it will be joined by
a second telescope, \PStwo, within the next two years
\citep{Wainscoat2013}. At the same time, the CSS G96 system will
increase the area of its field of view by a factor of $\sim4$, while
the CSS 703 site will increase the area of its field of view by a
factor of $\sim2.4$ \citep{christensen2012}.  

\subsubsection*{Asteroid Terrestial-impact Last Alert System (ATLAS)}

The ATLAS survey is expected to begin regular operations in early 2016
\citep{Tonry2011}.  They plan to survey the entire sky 4$\times$ each
night to $V\sim 20$ with relatively small but very wide-field
telescopes.  The system has one major limitation from the perspective
of detecting TCOs --- its relatively bright limiting magnitude --- but
this is compensated somewhat by its $\sim 4\arcsec$\,pixels that
dramatically reduce trailing losses.  The large pixel scale and
associated large astrometric uncertainties limits their ability to
detect all but the largest GCRs (see
fig.~\ref{fig.TCO_GCR_distribution}) but they expect to deliver moving
objects to the MPC within minutes of the final image acquired at a
boresite so that followup activity can begin almost immediately.
There are about $8.4\times10^{-3}$ TCOs at any time on the night sky
with $V<20.0$ and $\omega<15\arcdeg/\days$, a rate of motion
that will leave just 5 pixel long trails on the ATLAS images.  We thus
expect that ATLAS should detect about 0.1~TCOs/lunation or,
equivalently, about a 85\% chance of detecting a TCO in its first year
of operation.  One advantage to the bright limiting magnitude is that
almost all moving objects with $V<20$ will already exist in the MPC
catalog so that any new moving object that appears in the ATLAS survey
must be interesting --- \eg\ an NEO, asteroid cratering or disruption
event, or TCO.  While the ATLAS TCO detection rate might be relatively
low it is guaranteed that the discovered objects will be bright,
large, and relatively easy to track over many nights.

\subsubsection*{Space Surveillance Telescope (SST)}

We include in table~\ref{t.TCOSurveyPerformance} a row for the DoD's
Space Surveillance Telescope \citep[SST; \eg\ ][]{Monet2013} because
it will almost certainly become the leading asteroid discovery survey
if it is directed to search for NEOs and performs as expected.  It
is currently tasked with surveying the sky close to the equator for
artificial satellites but may eventually incorporate a sizable NEO
survey component much like the history of the Lincoln Near-Earth
Asteroid Research system \citep[LINEAR; \eg\ ][]{stokes2000}.  The
SST's short exposures limit TCO trailing losses and their all-sky
coverage to $V\sim22$ make $\sim0.4$ TCOs available to the system on
the sky plane at any time.  Like ATLAS, SST surveys the entire sky
every night so it is natural that there are $\sim50\times$ more TCOs
available to SST on the sky but because it surveys $\sim2$~magnitudes
fainter than ATLAS it `saturates' on the TCOs, \ie\ finds the
available ones, and the sky takes longer to `refresh' so that the TCO
discovery rate per day is only about double the ATLAS rate.

\subsubsection*{Large Synoptic Survey Telescope (LSST)}

The Large Synoptic Survey Telescope (LSST)
is nearly the ultimate TCO detection machine with its $\sim$9.6\,deg$^2$ FOV, $V\sim24.7$ limiting magnitude, and
15$\second$ exposure pairs, but it `only' surveys
about 20-25\% of the sky each night \citep{Ivezic2008}.  Its short and
back-to-back pairs of exposures should make identifying fast moving
objects easy and eliminate much of the confusion from systematics and
noise in the image plane.  Their $0.2\arcsec$ pixel scale in
combination with their excellent site on Cerro
Pach$\mathrm{\grave{o}}$n will mean that most TCO detections will be
trailed in their $35\second$ exposures (a 15$\second$ back-to-back
exposure pair with an intervening 5$\second$ read out time).  Our LSST
TCO detection estimate in table~\ref{t.TCOSurveyPerformance} used a
peak rate of detectable motion of $10\arcdeg/\days$ for which a TCO
would leave a 33\,pixel long trail in the system's focal plane.  We
expect that LSST could detect about 1.5~TCOs/lunation but a major
problem with this success rate is that most will be too faint for
followup by other observatories.  Another possible problem is that
with only 2 detections/night it will be difficult to measure the GCR.
Two possible ways to ameliorate the problem are to 1) treat the 2
exposures in a back-to-back pair as two separate detections which
would provide 4 detections per night or 2) it may be possible that the
trails in the two exposures have slightly different position angles
due to parallax.  The additional spatial information from the parallax
would allow for a significant reduction in the ephemeris
uncertainty. By the time LSST re-images the same TCO, if it is still
visible at all, the object will have moved a considerable distance on
the sky and manifest itself with an entirely different rate of motion,
position angle, and apparent brightness.

\subsubsection*{Subaru telescope with Hyper Suprime-Cam (HSC)}

A targeted survey with Hyper Suprime-Cam \citep[HSC;][]{Takada2010} on
the Subaru telescope located on Mauna Kea, Hawaii, could be very
effective at discovering TCOs.  The 870\,Mpix HSC with a $\sim
1.5\arcdeg$ wide FOV mounted on a 8.2\,m telescope at one of the best
sites in the world could detect small TCOs when they are well beyond
the Moon.  For our study we assumed that the HSC can reach $V=24.5$
for point sources in 15\,s exposures with sidereal
tracking.\footnote{Our assumption is based on the Subaru SuprimeCam
  exposure time calculator ({\tt
    http://www.naoj.org/cgi-bin/img\_etc.cgi}) for the $r'$ filter but
  is somewhat more optimistic because we assume that a minimoon survey
  would either use no filter or a wide-band filter.}  We restrict our
calculation to those TCOs with apparent rates of motion
$<1\arcdeg/\days$ so as to limit the effects of trailing of the
detections to less than a typical PSF of $\sim0.6\arcsec$.  With a
20\,s readout time this system could survey $\sim450$\,deg$^2$/night
centered on opposition and allow detection of TCOs of only 0.5\,m in
diameter at their mean geocentric distance of $\sim4.0$\,LD.
While there are $\sim$1.1~TCOs visible in the night sky with $V<24.5$
and moving slower than $1\arcdeg/\days$ the targeted HSC survey should
detect about 0.4~TCOs/night in the survey area or, equivalently, have
a $\sim$90\% chance of detecting a TCO in a 5-night observing run
spread over 40 days (because the TCO refresh time is about $8\days$ -
see table~\ref{t.TCOSurveyPerformance}).

\subsubsection*{Identifying TCOs by their great circle residual (GCR)}

Large GCR's relative to the astrometric uncertainty can provide better
short-arc ephemerides for the TCOs.
Figure~\ref{fig.TCO_GCR_distribution} shows that GCRs for the 4
different optical surveys decrease with increasing geocentric distance
as expected. A `plume' of TCOs with GCRs $>>1\arcsec$ is due to the
rare TCOs that are very close to Earth but
still moving slow enough to be detected by the survey.  For the ATLAS
survey the GCR is typically much less than the
astrometric uncertainty so that TCOs must be distinguished
by their fast rate of motion rather than GCR.  Most of the TCOs that
can be detected by the wide-field cameras with small pixel scales
will have significant GCRs
that should help in identifying them (assuming the detections
can be linked into tracklets) and also in calculating their ephemerides
for subsequent followup.  The lower-left panel in
fig.~\ref{fig.TCO_GCR_distribution} shows that most TCOs will not be
detectable with measurable GCR by LSST.  Our simulation assumed that
LSST would obtain 2 back-to-back exposures with $5\second$ time difference
and 60$\minute$ later obtain another back-to-back set.  We have
further assumed that the detections from each of the 4 exposures will
be identified rather than identifying only 2 detections in the stacked
back-to-back pairs.  While this cadence has many advantages to the
image processing pipeline it is not conducive to identifying TCOs.

\subsubsection*{Detectable TCO diameters}

The surveys discussed above are sensitive to TCOs from a few
$\mathrm{cm}$ to tens of $\mathrm{cm}$ diameter and the size of the
object at the peak in the number distribution decreases with the
aperture of the survey telescope (see
fig.~\ref{fig.TCO_detectable_size_distribution}).  The ATLAS survey
has a peak sensitivity to TCOs with $H\sim31$ or about 2\,m in
diameter.  The problem is that there are only 1-2 objects of this
diameter in orbit at any time and the probability that they pass near
opposition is relatively small (where the phase angle is close to zero
at a close enough distance to be detected by ATLAS).  \PSone\ could do
much better than ATLAS but is limited by the relatively small amount
of time devoted to surveying for moving objects and its inability to
identify trails in the images.  The wide area LSST and targeted HSC
surveys both have good sensitivity to TCOs of sub-meter diameter.  The
HSC survey discovers slightly larger TCOs than the LSST even though they have
roughly the same limiting magnitude because we have restricted the maximum
rate of motion to $<1\arcdeg/\days$, which means that the TCOs
must be more distant and therefore larger to be detectable.  This
advantage makes the TCOs easier to detect in dedicated followup
efforts with \eg\ radar or other optical telescopes.

\subsection{Meteor Survey Detection}
\label{ss.OpticalMeteorSurveyDetection}

The \citet{Granvik2012} NEO capture simulation yielded 18,096 TCOs of
which 189 or $\sim$1\% eventually struck Earth.  They normalize their
results to the \citet{Brown2002} bolide data and argue that about
0.1\% of meteors are TCOs prior to striking the atmosphere.

Considering that all-sky meteor surveys have existed for decades, and
that modern networks like ASGARD \citep[southern Ontario,
  Canada,][]{Brown2010}, CAMO \citep[Elginfield, Ontario,][]{Weryk2013},
and CAMS \citep[California, USA,][]{Jenniskens2011}, have the
capability of measuring the pre-meteor phase orbit, it might seem
surprising that are no reports that 1 in a 1,000
meteors were originally in geocentric orbit.  The solution to the puzzle is that not all meteors are equally
visible.  The apparent brightness of a meteor $\propto
m^{-2.02\pm0.15} s^{-7.17\pm0.41}$ where $m$ is the meteoroid's mass
and $s$ is its speed and we ignore an
insignificant dependence on the zenith angle that is consistent with
zero \citep{Sarma1985,Campbell2000}.  Thus, a meteor's apparent
brightness is exceedingly sensitive to its speed in the atmosphere.

\subsubsection*{Detecting TCO meteors with Cameras for Allsky Meteor Surveillance (CAMS)}

Meteoroids must have a mass $\ga0.6\g$ to be detected by CAMS at the
TCOs' sluggish $\sim11.2\kms$ geocentric impact
speed if we use a limiting $V$ magnitude of $+3.0$ \citep{Jenniskens2011}. This
suggests that TCOs must have a diameter of more than about $8\mm$ to
be detected assuming a typical meteorite grain density of
$3.0\g/\cm^3$ \citep{Britt2004}.

We calculate that from Oct 2010 through December 2011 CAMS observed
about 1.7~meteors/year with geocentric impact speeds consistent with
being in the TCO speed range of $[11.16,11.22]\kms$. The calculated
rate is the cumulative probability density of all detected meteors
with reported speeds within 3-$\sigma$ of the central value of
$11.19\kms$.  Extrapolating from the \citet{Brown2002} meteoroid SFD
down to $8\mm$ diameter suggests that there are
$(21\pm2)\times10^3$ TCOs of this size or larger striking Earth
every year corresponding to about $(4.1\pm0.4)\times10^{-5}$
TCOs/km$^2$/year.  The CAMS system monitors the sky above an altitude
of about 30$\arcdeg$ and at a typical meteor phase onset altitude of
about 75\,km the system therefore monitors about $3\times10^4\km^2$.
Accounting for the fact that the survey can only operate at night
(about $10\hourperday$) and for 30\% weather losses we estimate that
CAMs should detect about 0.4~TCO meteors/year.

The agreement to within a factor of about 4 between the predicted and
observed rate of meteors with TCO-like speeds is incredibly good
considering all the unknowns and uncertainties.  However, the
agreement should not be over interpreted, any value within a couple
orders of magnitude could have been argued as being due to detection
efficiency within the CAMS system, errors in the TCO model at the
smallest sizes due to Yarkovsky, contamination of the CAMS TCO-like
speeds by underestimated uncertainties or artificial satellite debris,
\etc

\subsubsection*{Detecting TCO meteors with the All Sky and Guided Automatic Realtime Detection system (ASGARD) and the Canadian Automated Meteor Observatory (CAMO)}

The ASGARD all-sky system has been in operation for about six years and recently
detected its first TCO-like meteor with an atmospheric entry speed of
$11.2\pm 0.8$\,km/s and mass of $\sim$100\,g (about 44\,mm diameter).
  Its speed is consistent with that expected for TCO meteors
but there is only a 10\% probability that it is actually a TCO-like speed
if we assume a Gaussian distribution in the speed uncertainty.  

We estimate that
ASGARD, with a limiting sensitivity of $V\sim-2$, should see about 0.01~TCOs/year of this size or larger implying
that its detection of a TCO-like meteor in only six years would be
unusual --- perhaps $\sim$15$\times$ higher than what we predict from
the TCO model.  On the other hand, with only a 10\% chance that the
ASGARD object has a TCO-like speed perhaps the numbers are in excellent
agreement.  Once again, we do not want to speculate too much on a
disagreement between the predicted and observed TCO meteor rate for
the same reasons as discussed above for CAMS.

Finally, the narrow-field influx cameras which are part of the CAMO
system have been in two-station operation for 3 years. In that time a
total of 685$\hour$ of observation have been manually reduced
resulting in 5,047 high quality double-station meteors.  An upper
limit to the system's collecting area at heights corresponding to an
ablation altitude appropriate to a 12$\km$/s entry speed (90$\km$) is
$\sim 350\km^2$.  At TCO geocentric
impact speeds of $\sim11.2\km$/s the limiting sensitivity of this
system is $V\sim+6$ corresponding to meteoroids with masses of $\sim
0.01\g$.  We identified 8 candidate TCO events with derived masses
between 0.01-0.12$\g$ (average$\,\sim0.04\g$) using the luminous
efficiency from \citet{Ceplecha1976} after detailed examination of all
5,047 meteors, eliminating those with poor geometry \citep{Musci2012},
and retaining only those with speeds (based on the average speed
during the first half of their visible trails) within 1$\sigma$ of
11.2$\km$/s . Manual examination of the solutions for the 8 events
shows that 3 have convincing dynamic flight solutions with geocentric
impact speeds at or below 11.2$\km$/s. The other five events show varying degrees of deceleration
more suggestive of true initial speeds above 12$\km$/s.

It is clear that it is difficult to determine accurate masses and speeds
of the meteors in the range of detectable TCOs but, for
comparison, a $0.04\g$ meteoroid has $H\sim45.5$. We predict that
there are about 650 TCO meteors per day over the Earth's entire
surface with $H<45.5$ and that CAMO should have seen about 0.01 of them given its observing time, sky coverage, weather losses, dark time,
\etc. The two orders of magnitude discrepancy between our prediction
for the number of TCO meteors that CAMO detects and their actual
results are in stark contrast to the factor of 4 under prediction for
CAMS (see above).  We attribute the differences to the difficulties in
measuring meteor speeds, and especially their masses, coupled with our
extrapolation of the TCO SFD to meteoroids that are almost ten orders
of magnitude smaller in mass than those that anchored the
\citet{Granvik2012} TCO SFD.


\section{Space-based Infrared Detection of TCOs}
\label{s.InfraredDetection}

\begin{figure}
\centering
\ifgrayscale
\hspace*{-0.512cm}
\includegraphics[scale = 0.45]{TCO_12micron_skyplane_residency_v2_bb_gray-eps-converted-to.pdf}
\hspace*{-0.512cm}
\includegraphics[scale = 0.445]{TCO_12micron_skyplane_residency_l1_0_5_meter_v2_bb_gray-eps-converted-to.pdf}
\else
\hspace*{-0.512cm}
\includegraphics[scale = 0.45]{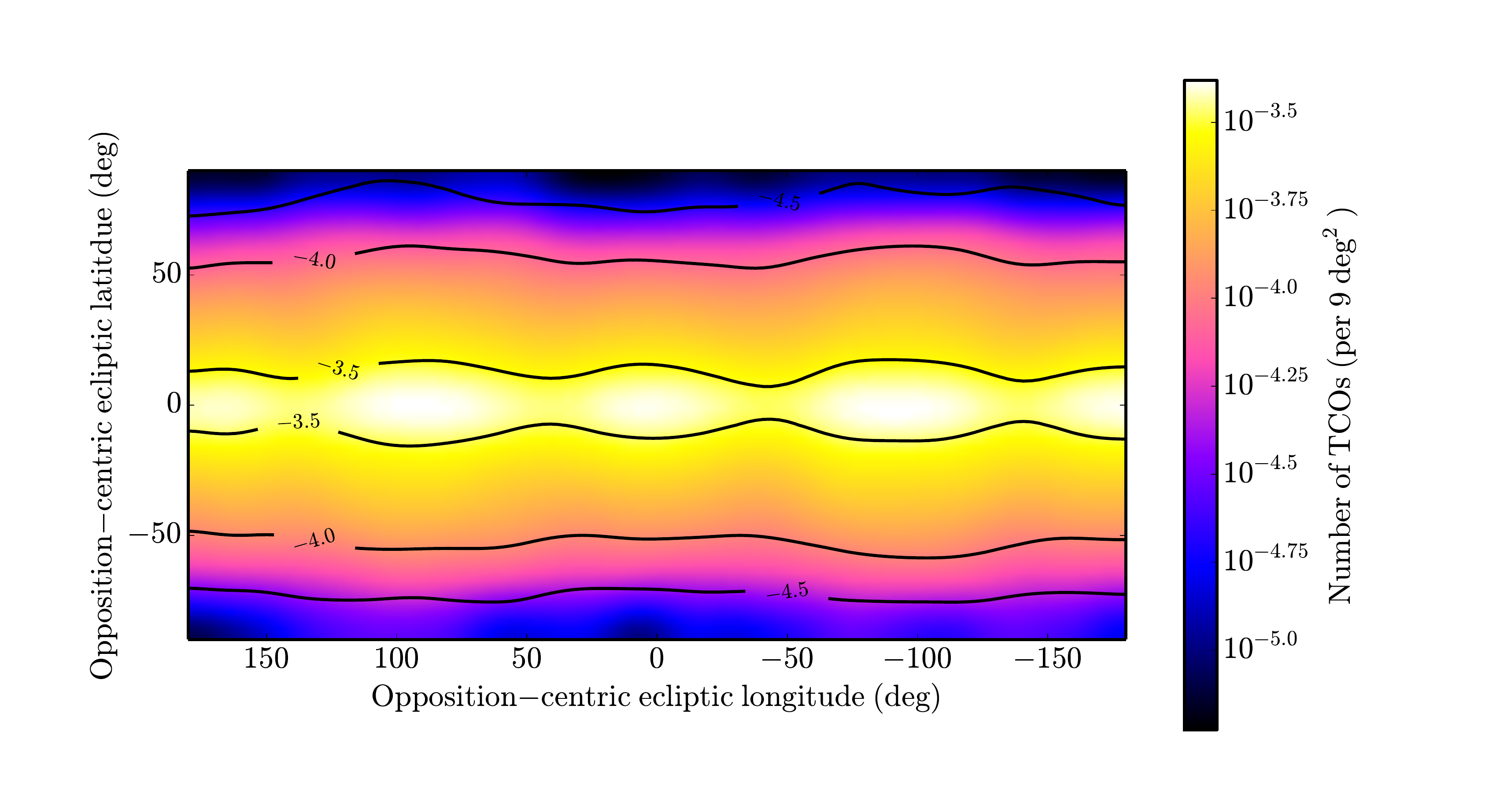}
\hspace*{-0.512cm}
\includegraphics[scale = 0.445]{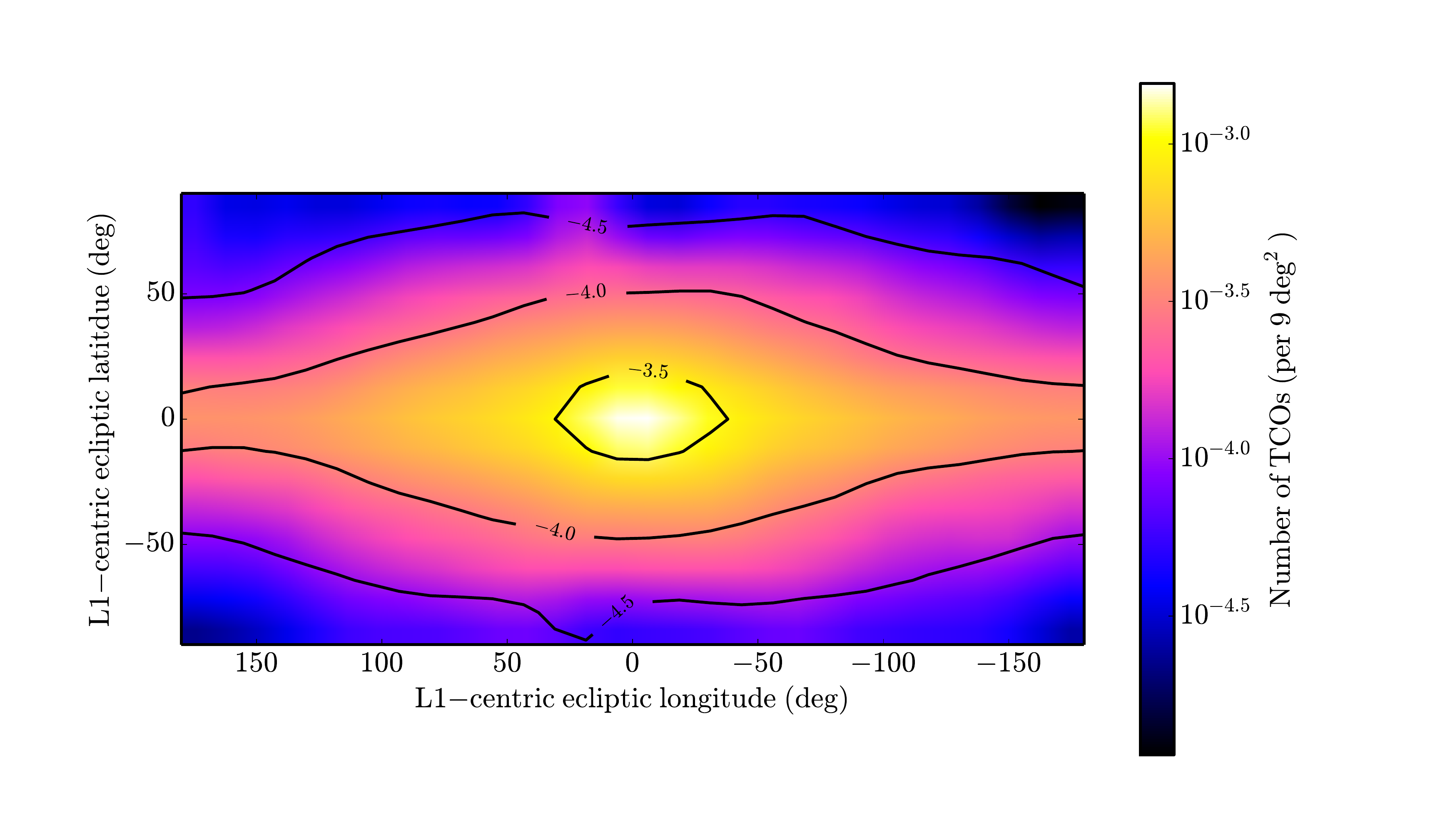}
\fi
\caption{Geocentric TCO sky plane number distribution in WISE's 12
  micron W3 band for TCOs with flux\,$>0.65$\,mJy from (top) geocenter
  and moving at $<3\arcdeg/\days$ and (bottom) from a spacecraft with
  a 0.5 meter aperture mirror at the Earth-Sun L1 Lagrange point capable of
  detecting TCOs moving at $<10\arcdeg/\days$.  The center of the
  figure is in the direction of Earth as viewed from L1.}
\label{fig.TCO_12micron_skyplane_residency}
\end{figure}

\begin{figure}
\centering
\hspace*{-1.1cm}
\ifgrayscale
\includegraphics[scale = 0.6]{TCO_GCR_optical-eps-converted-to.pdf}
\else
\includegraphics[scale = 0.6]{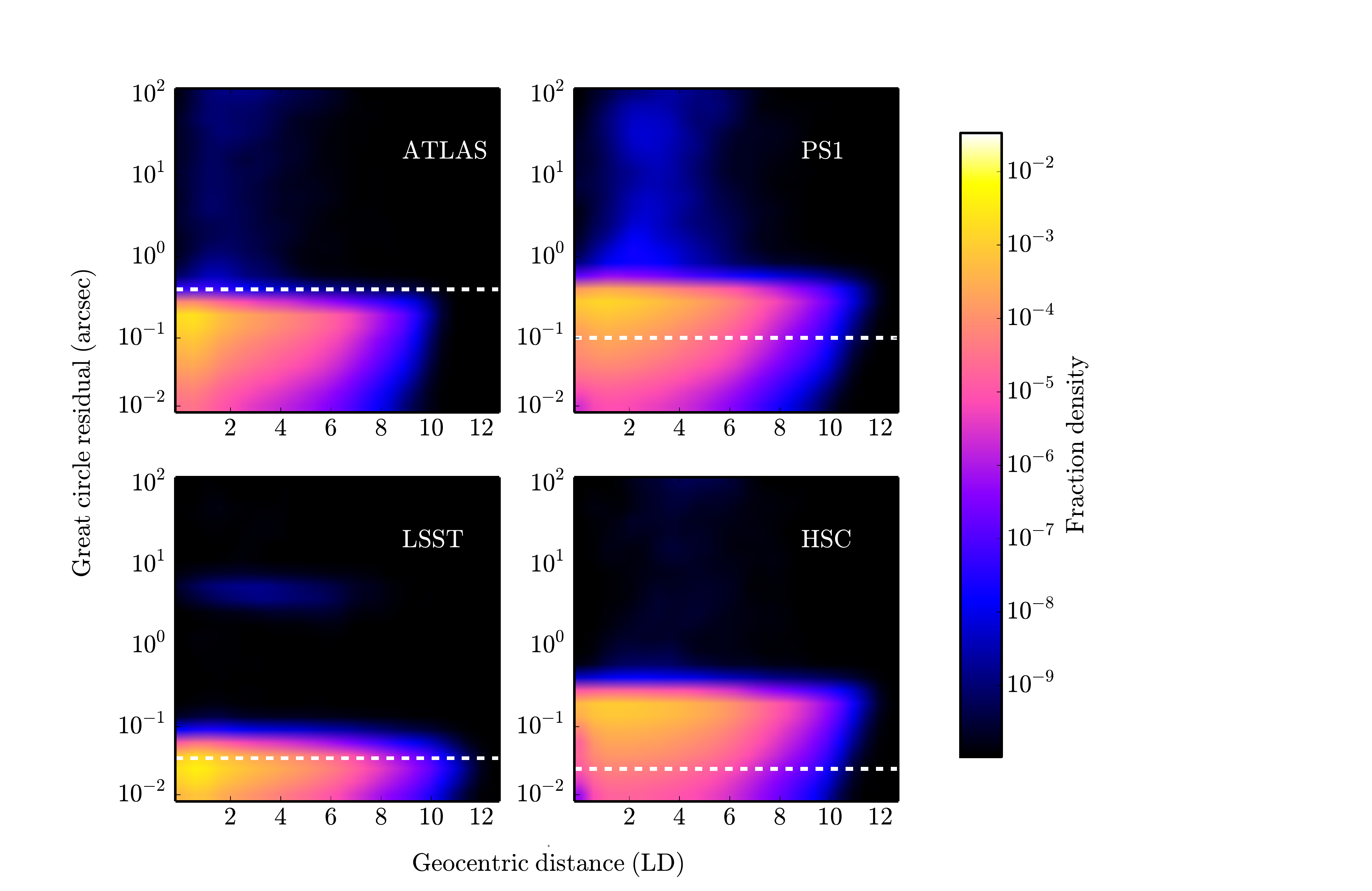}
\fi
\caption{True GCR versus geocentric distance for TCOs detectable with
  the ATLAS, PS1, HSC and LSST optical surveys.  The GCRs were
  calculated for the typical or expected number of exposures (3,4,3,2)
  and total time between first and last exposures of (40$\minute$,
  60$\minute$, 40$\minute$, 60$\minute$) for (ATLAS, PS1, HSC, LSST)
  respectively.  The dashed white lines provide a rough estimate for
  the astrometric uncertainty typical of each survey at $\sn=5$ given
  by the (pixel scale)/5. The number of objects in the GCR-$\Delta$
  range, $(\mathrm{GCR}\pm\dif\mathrm{GCR}/2,\Delta\pm\dif\Delta/2)$,
  is the product of the fraction density and
  $(\dif\mathrm{GCR},\dif\Delta)$.}
\label{fig.TCO_GCR_distribution}
\end{figure}

\begin{figure}
\centering
\ifgrayscale
\includegraphics[scale = 0.35]{TCO_optical_size_distribution-eps-converted-to.pdf}
\includegraphics[scale = 0.35]{TCO_IR_SFD-eps-converted-to.pdf}
\else
\includegraphics[scale = 0.35]{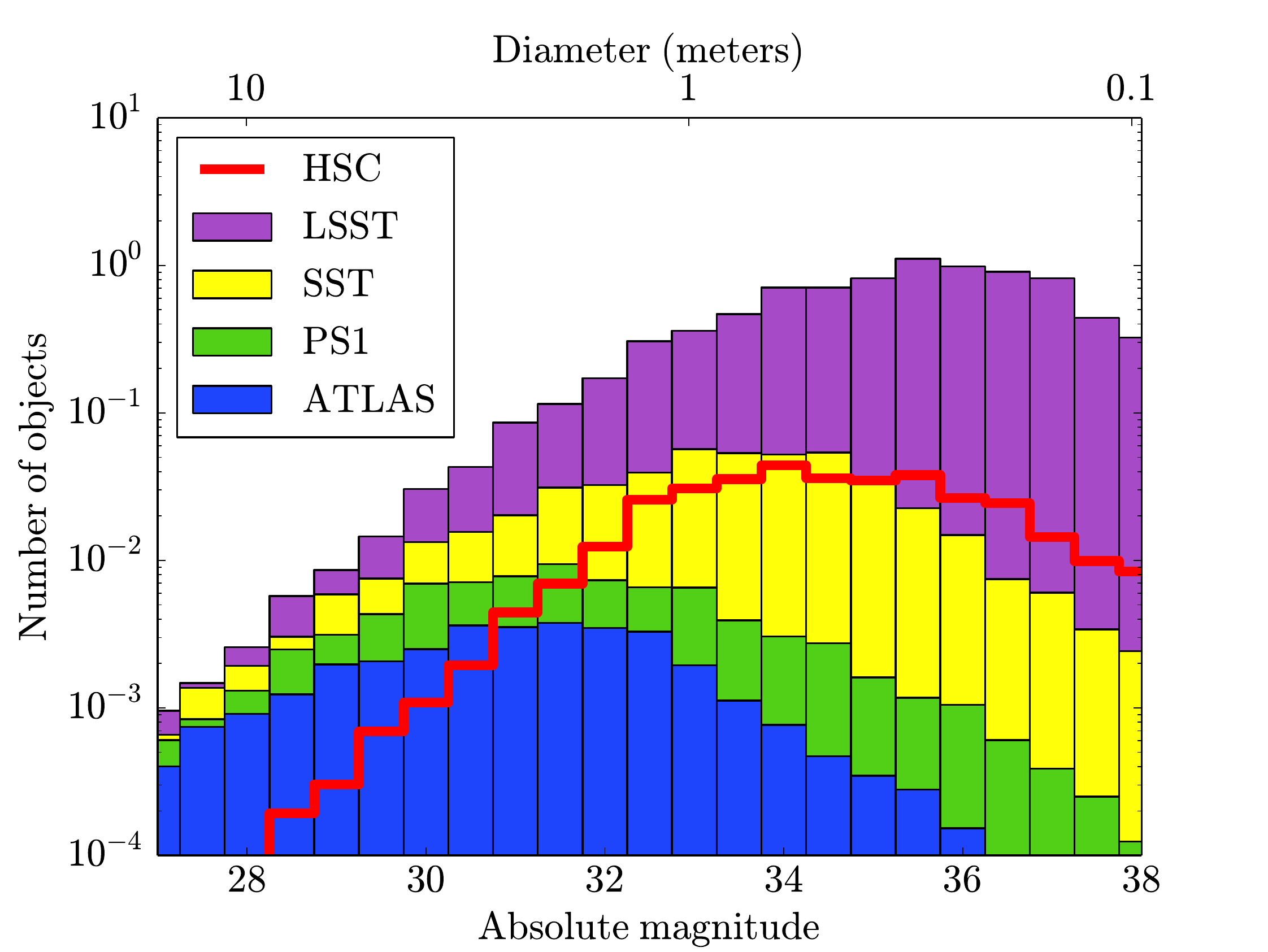}
\includegraphics[scale = 0.35]{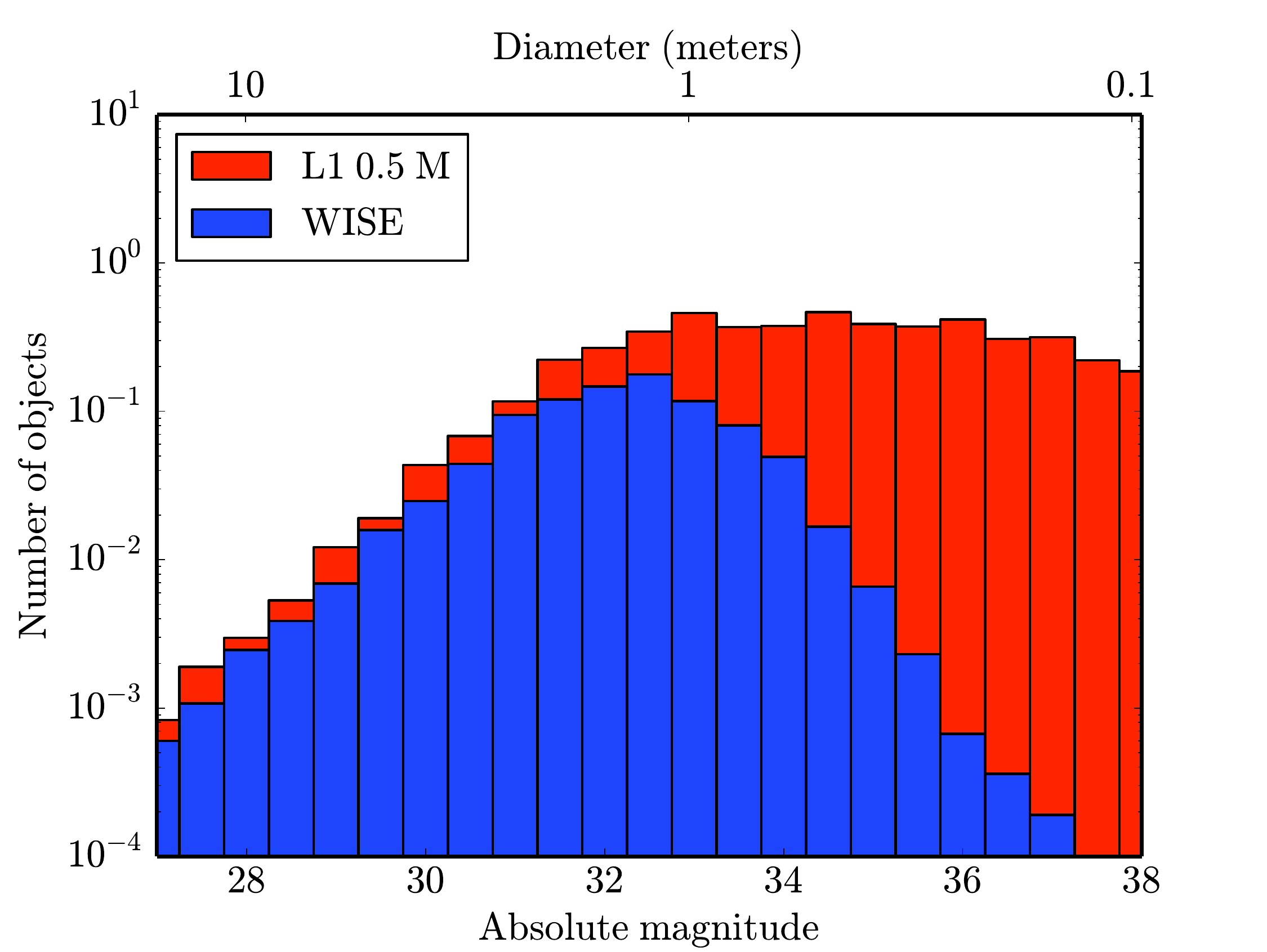}
\fi
\caption{Size distribution of detectable TCOs for (left) four different
  ground-based optical surveys as described in \S\ref{ss.OpticalTelescopeDetection} and (right) two different space-based IR surveys as described in \S\ref{s.InfraredDetection}.  Each bin represents the number of TCOs per
  0.5 absolute magnitude on the entire sky detectable by the survey.
  \ie\ it is not corrected for the survey's sky coverage or cadence.  Survey capabilities are provided in table~\ref{t.TCOSurveyPerformance}. }
\label{fig.TCO_detectable_size_distribution}
\end{figure}

Ground-based infrared facilities do not currently have the sensitivity
and wide-fields necessary for the TCO survey but the successful WISE
spacecraft mission \citep[\eg][]{Wright2010} and its asteroid-detecting NEOWISE sub-component \citep[\eg][]{Mainzer2011a} shows that
space-based IR surveys can be very effective (see
fig.~\ref{fig.TCO_12micron_skyplane_residency}).  (Our method of
calculating the IR flux from TCOs is outlined in
\ref{appendix.InfraredDetection}.)  Thus, we use the NEOWISE mission
as a baseline for an IR TCO survey even though future IR spacecraft
surveys such as the proposed NEOCam mission
\citep{McMurtry2013,Mainzer2006} will certainly be even more effective
at asteroid discovery.

To characterize the utility of a space-based TCO survey we estimate
the number of TCOs that might have been detected by NEOWISE in their
cryogenic 12$\mu$m W3 band images, the band most sensitive to
meter-scale TCOs as described in \S\ref{s.SteadyStateTCOPopulation}.
Images in this band were sensitive to sources with flux
$\Phi_{W3}>0.65$\,mJy \citep{Wright2010} and the fastest solar system
object they reported to the MPC was moving at $3.22\arcdeg/\days$.
The WISE FOV was 47$\arcmin$ with a 90\,minute re-visit time at each
sky location.  In the worst-case scenario a TCO moving at up to
$6\arcdeg/\days$ should still be detected 3$\times$ but in
practice NEOWISE required that the object be detected $>3\times$ to
reduce the false detection rate.  We find that there are $\sim 1$ TCOs
on the entire sky plane brighter than the specified flux limit and
moving slower than a slightly more conservative $3\arcdeg/\days$
(see fig.~\ref{fig.TCO_12micron_skyplane_residency}).  This
corresponds to about 0.005 TCOs in a 47$\arcmin$ wide strip extending
360$\arcdeg$ around the sky through the east and west quadratures ---
the region surveyed by WISE in a 90\,minute time interval.  With a
$\sim 2$\,day refresh time for TCOs with these properties we calculate
that WISE had about a 7\% probability of detecting one in each of the
10 months of the survey or about a 50\% chance of detecting one TCO
during the spacecraft's W3 band's operational lifetime.

WISE's peak sensitivity is for TCOs in the 1\,m size range ---
well-tuned to the expected size of the largest TCOs expected in the
steady-state (see fig.~\ref{fig.TCO_detectable_size_distribution}).
Since the WISE mission's cryogenic lifetime of about 10~months is well
matched to the average TCO lifetime it is likely that a WISE-like
survey would detect one of the 1-2 one-meter scale TCOs in orbit
around Earth at any time.  Larger objects are unlikely to be detected
simply because they are unlikely to be captured and the smaller
objects are not detected due to the flux limitations.  Thus, a
NEOWISE-like spacecraft mission could be an effective technique for
detecting the largest, and therefore most interesting, TCOs available
in the steady state.

NEOWISE is not known to have reported a TCO to the MPC despite our
calculation that it had a $\sim$50\% TCO detection efficiency.  There
are several possible explanations for the lack of a TCO detection
including, in no particular order of likelihood: 1) the probability of
NEOWISE {\it not} detecting a TCO was also 50\% 2) TCOs will have
higher GCRs than the NEOs that NEOWISE was tuned to detect and may
therefore not be detected with the same efficiency 3) ground-based
followup of NEOWISE's NEO candidates was required to confirm their NEO
status, but ground-based followup sites assume that the objects are on
heliocentric orbits when creating ephemeris predictions so that
recovery of objects that are actually on geocentric orbits will be
unlikely if not effectively impossible and, finally, 4) it is always
possible that our TCO model is incorrect.

WISE was not originally designed to identify asteroids and was not
optimized for TCO discovery.  This work concentrates on the
capabilities of existing assets in detecting TCOs but we have begun
studies to examine the capability of IR spacecraft optimized for TCO
discovery.  Our first tests assumed that 1) the spacecraft will be
located at the Sun-Earth L1 point so that the highest TCO sky plane
density and Earth are at opposition as viewed from the spacecraft (the
sky plane distribution is illustrated in
fig.~\ref{fig.TCO_12micron_skyplane_residency}) 2) mirror diameters of
$0.25\meter$, $0.5\meter$ and $1\meter$ and 3) the use of the WISE W3
passband with sensitivity from about $7.5\um$ to $16\um$ as
illustrated in fig.~\ref{fig.TCO_WISE_Flux}.  Under these assumptions
our calculations suggest that there are about 1, 5 and 30 TCOs
respectively on the sky plane {\it at any time} which suggests that this type of mission could provide
a steady stream of TCOs for ground-based physical characterization
(rotation state, colors and/or spectra) or even opportunities for
rendezvous, retrieval, or resource studies and utilization
\citep[\eg][]{Chyba2014}.  Actual TCO surveys in the IR from L1 would
need to avoid surveying too near the Earth and Moon but detailed
modeling of their performance is beyond the scope of this work even
though the prospect of TCO discovery with this type of mission is
promising.


\section{Radar Detection}
\label{s.radarDetection}

There are multiple radar systems capable of detecting TCOs either
while in orbit around Earth or as they plummet through the atmosphere.

There is little in the literature about detecting meteoroids but a
1994 DoD space surveillance network campaign \citep{Schwan1995}
identified 3 objects on similar orbits whose large semi-major axes,
high eccentricity and retrograde inclinations were surprising at the
time because the `origin of satellites in this region is uncertain'.
The 10 to 20$\cm$ diameter objects (larger if they were actually stony
asteroids) did not correspond to any known launch and the study
concluded that this `class of debris object may warrant further
evaluation'.  Their orbits matched the TCO orbit distribution very
well.  There is no doubt that the capabilities of the Air Force Space
Surveillance Systems have improved in the past twenty years and they
may now be capable of regularly discovering TCOs.

In the remainder of this section we focus on detecting TCOs with
existing mono-static and bi-static radar facilities (\eg\ Arecibo and
Green Bank) and meteor radar facilities that are available to the scientific community.

\subsection{Meteor Radar Detection}
\label{ss.MeteorRadarDetection}

Ground-based meteor radar systems efficiently detect the
ionization left behind in the atmosphere as a meteoroid
ablates.  They typically have high power and large aperture
(HPLA) because the radial scattering cross section is small. The radar echo is often referred to as
`head' echoes \citep{Baggaley2002}.

The radar detectability of 
meteors has been found empirically to be $\propto m^{0.92} s^{3.91}$
\citep{Verniani1973} where $m$ is the mass of the meteor in grams and
$s$ is the speed in $\kms$, but the dependence on speed becomes
much steeper at the small speeds typical of TCO meteors
 \citep{Weryk2013b}.  Our examination of $\sim$ 4 million speed
estimates made by the Canadian Meteor Orbit
Radar system \citep[CMOR;][]{Jones2005} found 1,790 with speeds below
$11.2\kms$.  Without additional information it is not
possible to determine if any were TCOs as opposed to
a decelerated, unbound meteoroid, as all of these detections were
below 85$\km$ altitude where the meteors have already decelerated due to atmospheric interactions at higher altitudes.

TCO head echo detection with HPLA is difficult because the small HPLA beam size
corresponds to a limited atmospheric collecting area of order a
  few to 10s of $\km^2$ or less \citep[\eg][]{Brown2001,Murray2004} but they
have the capability of detecting ablation masses larger than about
10$^{-8}\kg$ \citep{Ceplecha1998}.  Measured HPLA meteoroid speed
distributions show a strong peak at $\sim55\km$/s with only a few
detections at TCO impact speeds of $\sim11.2\kms$ \citep{Hunt2004}.
Our predicted HPLA TCO detection rate is about 8/year after
extrapolating down to these tiny masses with the nominal
\citet{Granvik2012} model.  \ie\ the HPLA systems should detect about
8 TCO meteors/year if they are 100\% efficient and operating
$24\hourperday$ and $365\dayperyear$.  The factor of 2-3$\times$ 
difference between the predicted TCO rate and the observed HPLA rate
is not significant given the uncertainty in the extrapolation to the small
TCO sizes and the reality that HPLA systems are not 100\% efficient
all the time.

\subsection{Direct meteoroid radar detection}
\label{ss.MeteoroidRADARDetection}

There have been long-standing efforts to detect orbital debris with
radar \citep{Stokely2009} but the only reported attempt to directly
discover meteoroids in space with radar was with the NORAD PARCS
system \citep{Kessler1980}. They identified 31 candidates in 8.4 hours
of observations with an average diameter of $\sim 7\cm$ ($H\sim39$),
comparable to the size of the TCOs in this study.  However, these
meteoroids were not gravitationally bound to Earth because their geocentric speeds were significantly higher than Earth's escape speed \ie\ they were not
TCOs.

Detecting known asteroids by their reflected radar signals is now
routine at the Arecibo and Goldstone facilities where over 450 Near
Earth Asteroids (NEA) have been detected.\footnote{as of May 2014 ---
  {\tt http://echo.jpl.nasa.gov/asteroids/}}  They bounce radar off
asteroids on a weekly basis and the smallest detected
objects to date are\footnote{{\tt http://echo.jpl.nasa.gov/{$\scriptstyle\sim$}lance/small.neas.html}} \XB\ and \RH\ (the latter being the only known TCO), both with $H \sim 30$
corresponding to 2-3$\meter$ in diameter.  However, they have not
discovered new asteroids other than the orbiting companions of the
targeted objects because the radar beam has limited angular coverage
and sensitivity that drops off like $\Delta^4$ where $\Delta$ is the
geocentric distance.

\begin{figure}
\centering
\hspace*{-1.65112cm}
\ifgrayscale
\includegraphics[scale = 0.55]{TCO_range_rate_vs_range_double-eps-converted-to.pdf}
\else
\includegraphics[scale = 0.55]{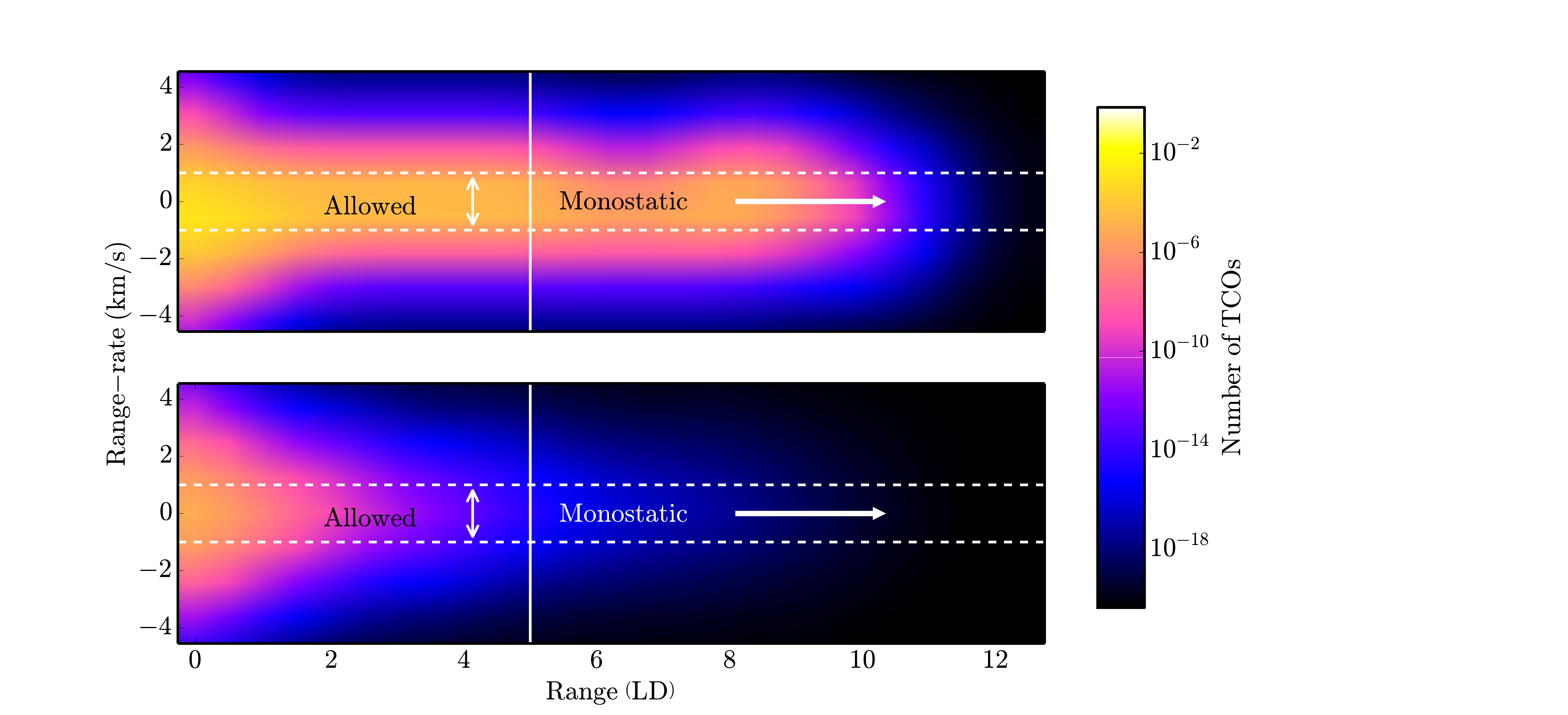}
\fi
\caption{Range-rate vs. range distribution for TCOs in a $1\arcdeg$
  opening angle cone centered on the east and west quadratures.  The
  top panel imposes no constraints on the objects' detectability while
  the bottom panel incorporates the constraints on $\sn \ge 14$ and
  range-acceleration $<10^{-4}\cm/\second$ (see
  fig.~\ref{fig.TCO-range-acceleration}).  The \sn\ requirement
  incorporates the effects of the TCOs' rotation rates.  The vertical
  solid line represents the minimum distance for which mono-static
  radar operations are possible with Arecibo.  We restrict the
  range-rate of detectable TCOs to $[-1\kms:+1\kms]$ as indicated by
  the region between the two horizontal dotted lines.}
\label{fig.TCO_rangeRate_vs_range}
\end{figure}

Figure~\ref{fig.TCO_rangeRate_vs_range} shows that the TCO's range and
range-rate distributions are within Arecibo's mono-static radar detection capabilities (in which
the site both transmits and receives the radar signal).  A
100\,cm/25\,cm diameter non-rotating TCO is detectable if it is within
$\sim$11$\,LD/\sim$6\,LD (we assume a minimum \sn=2.3 per second of
integration time as described in more detail below). The problem is that the reflected signal is doppler spread
by the object's rotation and at some size- and range-dependent
rotation rate the received \sn\ will be below the system's detection
limit.

The rotation rate distribution of meter-scale meteoroids is addressed
in \S\ref{appendix.rotationRates} where we describe our technique for
estimating the spin-rate distribution for TCOs of less than $10\meter$
diameter.  While the predicted median spin rates for objects of the
size we expect to identify with radar render them undetectable by
current radar assets there is a long tail of slow-rotating meteoroids
that can be detected.

Our calculations (see \S\ref{appendix.RadarSurveys} for an overview of
calculating radar detectability of asteroids) suggest that detecting
TCOs is essentially impossible with mono-static operations at Arecibo
because their minimum detectable geocentric distance of 4-5\,LD is
fixed by the time to switch between transmitting and receiving and
there are simply not enough TCOs rotating slow enough beyond that
distance to make the technique viable (see
fig.~\ref{fig.TCO_rangeRate_vs_range}). The Goldstone array has the
capability of observing objects as close as 2\,LD, but the increase in
signal from the decrease in the minimum observable distance is not
enough to overcome the several factors decrease in sensitivity
compared to the Arecibo array.

\subsubsection*{Detecting TCOs with bi-static radar}

There may be opportunities to discover TCOs using bi-static operations
where the radar signal is transmitted by Arecibo and received at Green
Bank \citep{Benner2002}.  Routine radar detection of known asteroids
requires that the object's topocentric range and range-rate be known
so that the returned signal is sampled in the appropriate range and
range-rate bins.  A search for unknown TCOs requires identifying the
reflected signal in range-rate phase space (see
fig.~\ref{fig.TCO_rangeRate_vs_range}) and to do so requires limiting
the range-rate extent and binning the returned signal.  

\begin{figure}
\centering
\hspace*{-.25112cm}
\ifgrayscale
\includegraphics[scale = 0.55]{TCO_drifting-eps-converted-to.pdf}
\else
\includegraphics[scale = 0.55]{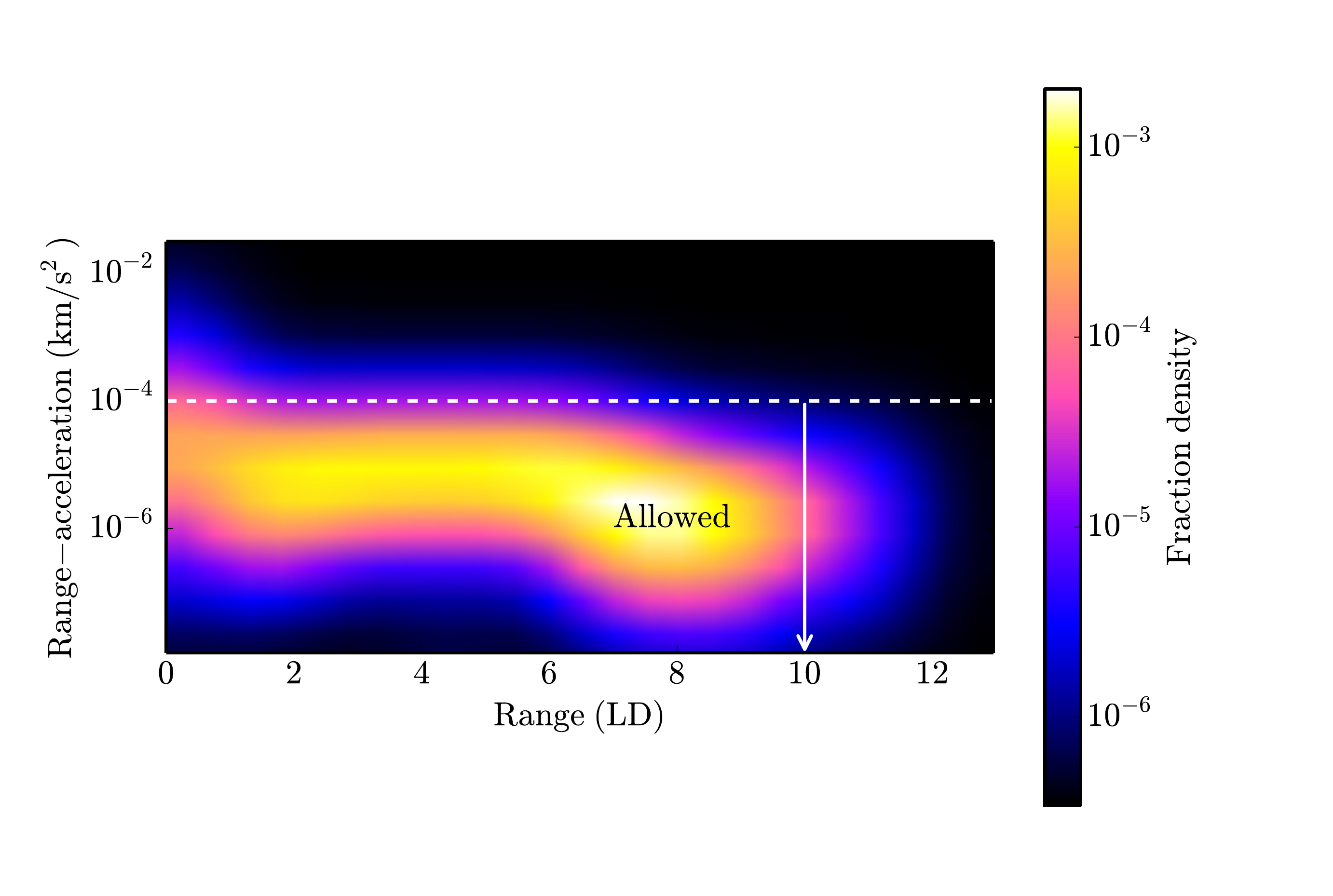}
\fi
\caption{Range-acceleration ($\ddot\Delta$) vs range distribution for
  minimoons near the east and west quadratures. There is a steep rise
  in range-acceleration for minimoons at ranges of $<1$\,LD. More than
  99\% of TCOs are accelerating at $<10^{-4}\kms^{-2}$ as indicated by
  the dashed horizontal line. The number of objects in the
  $\ddot\Delta$-$\Delta$ range,
  $(\ddot\Delta\pm\dif\ddot\Delta/2,\Delta\pm\dif\Delta/2)$, is the
  product of the fraction density and $(\dif\ddot\Delta,\dif\Delta)$.}
\label{fig.TCO-range-acceleration}
\end{figure}

For our
simulations we restrict the search to $-1\km/\second < \dot\Delta <
+1\km/\second$ because $\sim96$\% of the detectable TCOs are found
within this range.  The range-rate dimension was binned by
$10\cm/\second$ because $>99$\% of all TCOs have accelerations of
$<10\cm^2/\second$ (\ie\ not just the detectable ones; see
fig.~\ref{fig.TCO-range-acceleration}) and we will assume $1\second$
integration times so that a TCO remains in one bin during each
transmission.  Our simulation also assumes that each telescope
pointing direction will be scanned with up to 12 one-second
transmissions.  At the maximum allowed acceleration the TCO's
range-rate may be in a different bin in each scan, reducing the
\sn\ in an individual scan by a factor of about $3.5\sim\sqrt{12}$.
Our experience with radar asteroid detection suggests that the minimum
\sn\ is $\sim 2.3$ so that we require the total received $\sn=8$ for
  a TCO detection.  We used a realistic transmitting power of 0.9\,MW
  and a conservative radar albedo of 0.15 typical for S-type
  NEOs\footnote{The average of many published values summarized at
    \tt{http://echo.jpl.nasa.gov/$\scriptstyle\sim$lance/asteroid$\_$radar$\_$properties/nea.radaralbedo.html}.}
and calculated the returned \sn\ accounting for the TCO's rotation
rate as described in
\ref{appendix.RadarSurveys}. Each radar pointing direction has an
$\sim2\arcmin$ beam width in which we calculate there are $\sim4\times
10^{-6}$ detectable TCOs at any time at opposition and $\sim3\times
10^{-5}$ at the quadratures.  These values correspond to about
$5\times 10^{-3}$ and $3\times 10^{-2}$ detectable TCOs/deg$^2$ at the
two locations respectively.

The detectable TCOs are dominated by objects with $\Delta\la0.5$\,LD
and $-1\kms < \dot\Delta < +1\kms$
(fig.~\ref{fig.TCO_rangeRate_vs_range}) and must be objects in the
tail of the rotation period distribution with long rotation periods
(on the order of minutes or more).  Since the objects are so close to
Earth their apparent transverse speeds are high and the refresh rate
of objects in the radar beam is about $10\minute$ towards opposition
and $30\minute$ towards the east and west quadratures.  However, since
the TCO residence time is enhanced towards the east and west
quadratures at distances less than 1\,LD
(fig.~\ref{fig.TCO_rtheta_residency_noconstraints}), the probability
of discovering a TCO towards the east and west quadratures is larger
than towards opposition.  The east or the west quadrature is
always observable by Arecibo for at least 7-8 hours every day despite
the pointing limitations of the array.

\begin{figure}
\centering
\hspace*{-0.112cm}
\ifgrayscale
\includegraphics[scale = 0.57]{TCO_RADAR_SFD-eps-converted-to.pdf}
\else
\includegraphics[scale = 0.57]{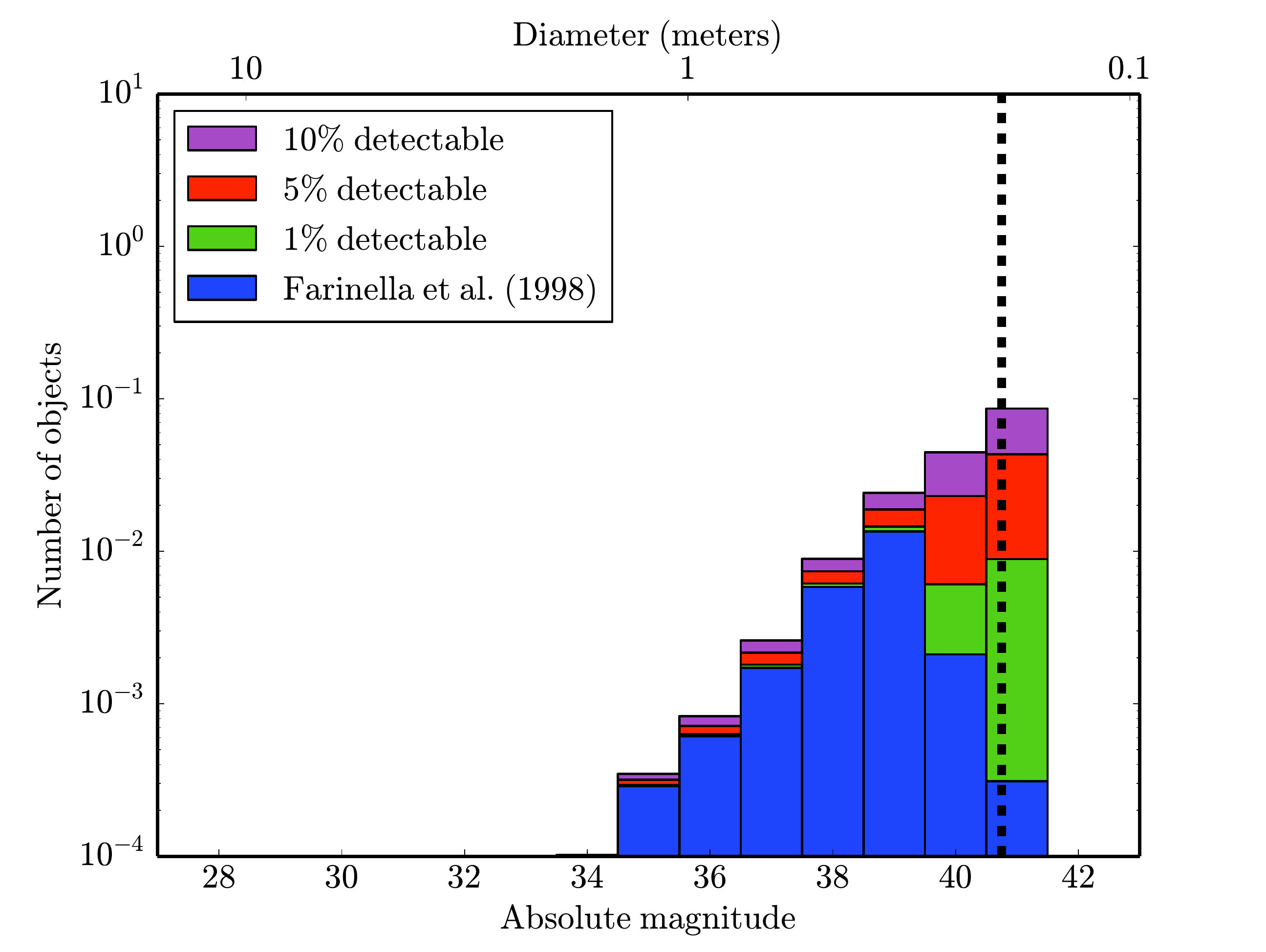}
\fi
\caption{Size distribution for TCOs discovered in our simulations of
  bi-static radar operations with the Arecibo and Green Bank
  facilities.  The four histograms represent different underlying
  rotation period distributions.  The nominal distribution of
  \citet{Farinella1998}, and 3 modified distributions that
  artificially increase the percentage of objects with detectable
  rotation periods to 1\%, 5\% and 10\%.  A dashed vertical line at
  $0.03\meter$ indicates where Rayleigh scattering prevents detections
  of small TCOs. \ie\ all detections in the simulation with diameter
  $\le3\cm$ have been removed but we have not eliminated any
  detections with diameter $>3\cm$.}
\label{fig.TCO_RADAR_SFD}
\end{figure}

Arecibo has a 20\% probability of detecting a TCO in 40$\hour$ of
observing in the east and/or west quadratures when we use our
conservative assumptions about the TCO rotation rate distribution as
described in \S\ref{appendix.rotationRates}. However, we have
empirical evidence that suggests the rotation rate distribution is
non-Maxwellian since it does not provide a good fit to rotation rates
of small asteroids in the Light Curve Database
\citep[LCDB,][]{Warner2009}.  The actual rotation rate distribution suggests
that $\sim$10\% of asteroids with $H>25$ have rotation rates that are
slow enough to have a negligible effect on their radar detectability
at geocentric ranges within a few LD. The probability of detecting a
TCO with 40$\hour$ of observing time increases to $30\%$, $60\%$ and
$80\%$ if we assume that $1\%$, $5\%$ and $10\%$ of TCOs respectively
are rotating slow enough to be detectable as suggested by the LCDB
data (fig.~\ref{fig.TCO_RADAR_SFD}). (All our
radar results impose a lower diameter limit of $3\cm$ because Rayleigh scattering starts to have a significant
effect on the detection of objects of about $3\cm$ diameter because
the radar wavelength is $13.6\cm$ \citep{Knott2004}.)

\begin{figure}
\centering
\hspace*{-0.6512cm}
\ifgrayscale
\includegraphics[scale = 0.57]{TCO_radar_orbits_all-eps-converted-to.pdf}
\else
\includegraphics[scale = 0.57]{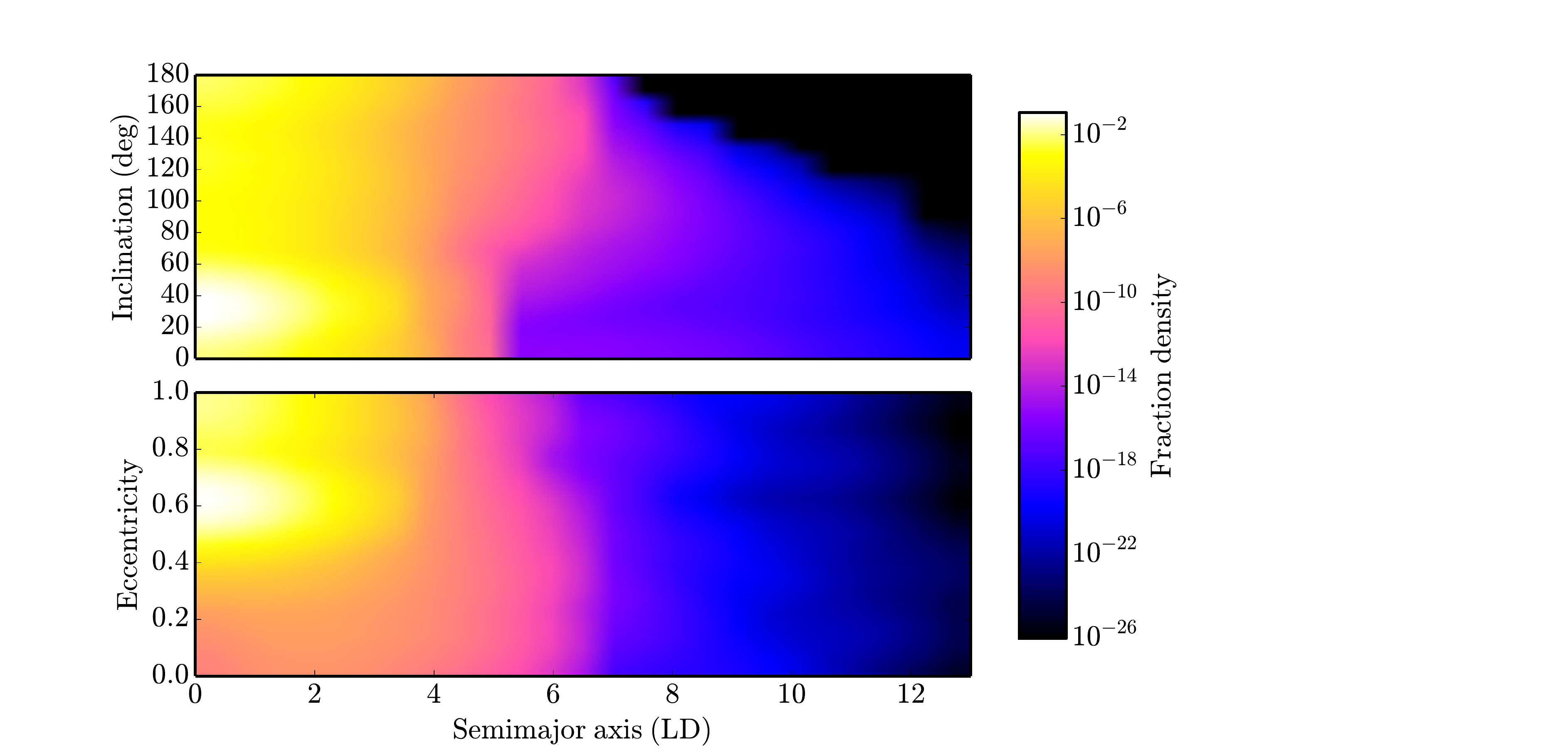}
\fi
\caption{Geocentric semi-major axis vs. inclination (top) and
  eccentricity (bottom) distributions for minimoons discovered in our
  bi-static radar simulations in the direction of the quadratures.  The
  distributions are dominated by TCOs with lifetimes $>2,000\days$
   with
  inclinations of $\sim 40^{\circ}$ and moderate eccentricities of
  $\sim 0.6$.  Letting $x$ represent either inclination or
  eccentricity, the number of objects in the $x$-$a$ range, $(x\pm\dif
  x/2,a\pm\dif a/2)$, is the product of the fraction density and
  $(\dif x,\dif a)$.}
\label{fig.TCO_radar_orbits_all}
\end{figure}

Our results suggest that radar-detected TCOs have a very different
orbit distribution from the entire steady-state population.  Fully
85\% of the radar-detected TCOs at the quadratures have TCO lifetimes
of $>2,000\days$ compared to just $\sim0.3$\% of the entire
steady-state population \citep{Granvik2012}.  The orbital element
distribution of all radar-detectable TCOs compared to the orbital
element distribution for radar-detectable TCOs with lifetimes in
excess of 2,000$\days$ tend to have smaller semimajor axes
(fig.~\ref{fig.TCO_radar_orbits_all}).  The preferential detection of
the long-lived TCOs has implications for spacecraft TCO rendezvous
missions because they have the advantage of being accessible by
spacecraft for years after their discovery \citep{Chyba2014}.


\section{Discussion}
\label{s.Discussion}

We have shown above that it will be challenging but not impossible to discover TCOs on a
regular basis with existing facilities but the scientific and exploration
opportunities might generate enough interest
to dedicate resources to the task.

One TCO discovery opportunity that was not explored above is
data mining of the Minor Planet Center's (MPC) so
called one-night-stand (ONS) file.  This list of detections that have
never been linked to known heliocentric objects with longer arcs could
be searched specifically for geocentric objects.  As mentioned in the
introduction, we have anecdotal evidence suggesting that unknown
geocentric objects were/are discovered and followed by asteroid
surveys but they are not flagged as interesting by the MPC because it
is difficult or impossible for them to distinguish these objects from
operational spacecraft or space junk.  However, the work of
\citet{Granvik2012} provides a means of comparing the derived
geocentric orbit to those expected for TCOs which are typically very
different from artificial satellite orbits.  Given the short arc
lengths and lack of spectroscopic or even colors of these historical
ONS detections it is unlikely that any TCO could ever be confirmed, but
if there were enough detected objects it might be possible to compare
the distribution of their geocentric orbits with the expected TCO
distribution.

Of course, there will always remain the possibility that the detected
objects were classified satellites because the types of orbits
occupied by TCOs are also the types of orbits that might be desirable
for `hiding' spacecraft --- objects on these orbits would spend a
significant fraction of their time far from Earth where they are very faint, and when they
are brighter and closer to Earth they would be moving extremely fast.  Either way, the objects are difficult to detect.

This work has addressed only the issue of TCO discoverability and
ignored issues related to observation cadence, followup, and time of
discovery relative to time of capture --- all important issues for
determining the viability of observation programs that would target
the TCO for physical characterization or spacecraft missions that
might attempt to intercept or even retrieve the object
\citep{Chyba2014}.  This kind of work would require a high fidelity
simulation of an asteroid survey system capable of integrating the
TCOs within the Earth-Moon system.

Even if TCOs are discovered by operational surveys it is important to
understand whether their orbits can be determined quickly and well
enough to allow followup or a spacecraft mission.  As is well
known from NEO followup efforts, short-arc extrapolation of discovery
tracklets for `simple' heliocentric NEO motion can quickly lead to
ephemeris sky plane uncertainties of many deg$^2$ making recovery
impossible.  We have begun to study the evolution of the orbital
uncertainties and error as a function of a TCO's observational arc and
find that the TCOs' orbits converge rapidly due to their proximity and
the advantage afforded the orbit determination by their topocentric
parallax \citep{Granvik2013b}.  It is our expectation that just a
dozen TCO observations over a few days should allow the orbit to be
known well enough to enable radar detections which will then
dramatically refine the orbit.

TCOs discovered with optical and radar assets may be distinguished
from space debris by their geocentric orbital elements. The vast
majority of space debris have circular geocentric orbits within 300 km
($8 \times 10^{-4}$ LD) of the Earth's surface \citep{Tingay2013}
where as TCOs which come within such a close distance have highly
eccentric orbits (fig.~\ref{fig.TCO_radar_orbits_all}).  This
comparison could rule out artificial from natural satellites as soon
as the orbital solutions of a TCO candidate were determined. 

TCOs may also be distinguished from space debris by their bulk density
that could be measured by examining how a candidate TCO is affected by
radiation pressure \cite[\eg\ ][]{Micheli2013}.  Indeed, the trajectory
 of \RH\ could not be reproduced without accounting for the effects of radiation pressure acting on the few-meter diameter object.  Precision ephemeris measurements with ground-based optical telescopes over the course of days or weeks could measure the radiation pressure
on the small TCOs
(personal communication with Marco Micheli, 2013), or sooner with
repeated radar observations given their superior range and range-rate
determination.  We expect that TCOs will have bulk densities similiar
to porous rock \citep[$1\g/\cm^{-3}$;][]{Britt2004} but artificial
objects on the meter-scale will have much smaller values \eg\ the
Apollo 12 fourth stage was recaptured in the Earth-Moon system on a
TCO-like orbit but its bulk density was $\sim0.02\g/\cm^{-3}$

Finally, compositional information obtained with radar or
spectroscopic follow up observations could be used to rule out or
determine artificiality of TCO candidates. Such was the case with the
object designated as J002E3 that was conclusively linked with the
Apollo 12 fourth
stage\footnote{\tt{http://www.jpl.nasa.gov/releases/2002/release$\_$2002$\_$178.cfm}}.


\section{Conclusions}
\label{s.Conclusions}

The detection of Earth's temporarily captured orbiting (TCO) asteroids
is challenging due to the small number of objects in the population in the steady state, their small sizes (the largest being only
on the order of 1 to 2$\meter$ diameter), their
large geocentric semi-major axes, and highly eccentric orbits.
Despite these challenges we find that a space-based IR survey system
could be effective at discovering the TCOs.  There may be up to a few
dozen TCOs detectable on the sky plane at any time in the IR with a
moderate size mirror (0.5 to 1.0$\meter$) on a spacecraft located near
the Earth-Sun L1 point (\S\ref{s.InfraredDetection}).  This kind of
survey system might be justified from the standpoint of identifying
objects that could be stepping stones to learning how to exploit
near-Earth natural resources \eg\ developing techniques for
navigation, interaction with, and mining small asteroids.  A targeted
ground-based TCO survey with a wide-field camera on a large telescope
could be effective for testing the \citet{Granvik2012} model \eg\ our
calculations suggest that a TCO survey with Hyper Suprime-Cam on the
Subaru telescope (\S\ref{ss.OpticalTelescopeDetection}) has about a
90\% chance of detecting a TCO in a dedicated 5-night survey.  While
this survey is useful for testing the principle it would not be an
effective means of guaranteeing a steady stream of TCOs for scientific
followup or rendezvous and retrieval missions.  Finally, it is
possible that TCOs may be directly discoverable with bi-static radar
(\S\ref{ss.MeteoroidRADARDetection}) but this possibility relies
on the rotation rate distribution being strongly non-Maxwellian for
the small TCOs. The long-lived TCOs discovered in this way would be valuable targets for small asteroid characterization
and rendezvous or retrieval missions.


\acknowledgments

\section*{Acknowledgments}

This work was supported in part by NASA NEOO grant NNXO8AR22G. PGB
acknowledges funding support from co-operative agreement NNX11AB76A and the
Canadian Natural Sciences and Engineering Research Council. MG was
funded by grant \#137853 from the Academy of Finland.  MCN and ESH
were supported under NEOO grant NNX12AF24G.


\clearpage

\appendix

\section{Appendix}

\subsection{Sky plane residence time distribution}
\label{appendix.SkyPlaneResidenceTimeDistribution}

We define the sky plane residence time $T_j$ for a particle $j$ as the
time it spends near ecliptic longitude $\lambda$ and latitude $\beta$
in the intervals $[\lambda-\Delta\lambda/2, \lambda+\Delta\lambda/2]$
and $[\beta-\Delta\beta/2, \beta+\Delta\beta/2]$.  We can write the
infinitesimal residence time in terms of the residence time density
$\rho_j(\lambda,\beta)$ such that $\dif T_j(\lambda,\beta) =
\rho_j(\lambda,\beta) \; \dif\lambda, \dif\beta$ and
\begin{equation} 
\label{eq.residence}
  \rho_j(\lambda,\beta)
    = \int\limits_{-\infty}^{+\infty} dt
       \; \delta( \lambda_{j}(t)-\lambda )
       \; \delta( \beta_{j}(t)-\beta )
\end{equation}
\noindent where we introduce the Dirac $\delta$-function and
$\lambda_j(t)$ and $\beta_j(t)$ represent the ecliptic longitude and
latitude of the particle at time $t$ respectively. The
residence time of the particle within the extended
range $\lambda_1 \le \lambda < \lambda_2$ and $\beta_1 \le \beta <
\beta_2$ is then
\begin{equation} 
 T_j(\lambda_1,\lambda_2,\beta_1,\beta_2)
  = \int\limits^{\lambda_2}_{\lambda_1} \dif\lambda \;
    \int\limits^{\beta_2}_{\beta_1} \dif\beta \; 
    \rho_j(\lambda,\beta).
\end{equation}

The sky plane residence time density for a population of
particles is the sum of the individual sky plane residence time
densities
\begin{equation} 
\label{eq.sum_residence}
  \rho(\lambda,\beta) = \sum_j \rho_j(\lambda,\beta)
\end{equation}

Letting the normalization constant
\begin{equation} 
\label{eq.constant}
  C = \int \dif\lambda \int \dif\beta \;\; \rho(\lambda,\beta)
\end{equation}
\noindent \ie\ the cumulative time spent over the entire sky by all
particles, the normalized residence time density is
\begin{equation} 
\label{eq.normalized_residence}
  \rho_N(\lambda,\beta) = \frac{1}{C} \; \rho(\lambda,\beta).
\end{equation}
Thus, $\rho_N(\lambda,\beta) \Delta\lambda \Delta\beta$ is the
fraction of time that all the particles spend within
$(\Delta\lambda/2, \Delta\beta/2)$ of $(\lambda, \beta)$.

\subsection{Sky plane number density}
\label{appendix.SkyPlaneNumberDensity}

Let the particles' cumulative $H$-frequency distribution (HFD) be
\begin{equation} 
\label{eq.HFD}
N_{HFD}(H) = 10^{\alpha(H-H_1)}
\end{equation}
\noindent \ie \ $N_{HFD}(H)$ is the number of particles with absolute
magnitude $< H$ and $H_1$ is therefore the absolute magnitude at which there is only one object in the population with $H<H_1$.

The differential HFD is then
\begin{equation} 
\label{eq.diff_HFD}
n_{HFD}(H) \; dH = \alpha \; \ln{10} \; 10^{\alpha(H-H_1)} \; dH
\end{equation}
\noindent \ie \ there are $n_{HFD}(H) \; dH$ objects in the interval
$dH$ at magnitude $H$.

Since $\rho_N(\lambda,\beta)$ is the normalized sky plane residence
time density, the differential sky plane number density of objects at
absolute magnitude $H$ is
\begin{equation} \label{eq.diff_number}
n(\lambda,\beta,H) \;  = n_{HFD}(H) \; \rho_N(\lambda,\beta).
\end{equation}
\noindent \ie\ the number of objects in an absolute magnitude interval
of $[H-\Delta H/2, H+\Delta H/2]$  around $H$ and in longitude $\lambda$ and latitude $\beta$ in the intervals
$[\lambda-\Delta\lambda/2, \lambda+\Delta\lambda/2]$ and
$[\beta-\Delta\beta/2, \beta+\Delta\beta/2]$ is
$n(\lambda,\beta,H) \; \Delta\lambda \; \Delta\beta \; \Delta H$.

The cumulative sky plane number density of objects brighter than $H_0$
at $(\lambda,\beta)$ is
\begin{eqnarray}
\label{eq.number}
\nonumber
N(\lambda,\beta,H_0) 
  &=& \int^{H_0} \; dH \; n(\lambda,\beta,H) \\ 
  &=& N_{HFD}(H_0) \; \rho_N(\lambda,\beta) 
\end{eqnarray}
so that the number of particles with $H<H_0$ in longitude and latitude
intervals of widths $(\Delta\lambda,\Delta\beta)$ around
$(\lambda,\beta)$ is $N(\lambda,\beta,H_0) \; \Delta\lambda
\; \Delta\beta$.

Thus, the number of particles with absolute magnitude $<H$ in the
range $\lambda_1 \le \lambda < \lambda_2$ and $\beta_1 \le \beta <
\beta_2$ is
\begin{equation} 
  \int\limits^{\lambda_2}_{\lambda_1} \dif\lambda \; 
  \int\limits^{\beta_2}_{\beta_1} \dif\beta \;
  N(\lambda,\beta,H).
\end{equation}

\subsection{Optical Detection of TCOs}
\label{appendix.OpticalDetection}

Following the nomenclature of \S\ref{appendix.SkyPlaneResidenceTimeDistribution}, the sky plane residence time density for an individual particle $j$ with absolute magnitude $H_0$ while it has apparent magnitude
$V(H_0)<V_0$ and rate of motion $\omega<\omega_0$ is identical to eq.~\ref{eq.residence} except for a
delimiter that constrains the rate of motion and apparent magnitude:
\begin{equation}
\label{eq.time_vis}
  \rho_j^{vis}(\lambda,\beta,V_0,\omega_0,H_0) \;
   = \int\limits_{-\infty}^{+\infty} dt \;
     \delta( \lambda_{j}(t)-\lambda) \;
     \delta( \beta_{j}(t)  -\beta ) 
        \Bigg|_{V(H_0)\;<\;V_0, \ \omega\;<\;\omega_0}
\end{equation}
The apparent magnitude $V$ is calculated with the techniques described in \cite{Muinonen2010} and the rate of motion $\omega$ is determined by the plane of sky motion.

If all the particles have absolute magnitude $H_0$ their sky plane
residence time density while they have $V(H_0)<V_0$ and
$\omega<\omega_0$ is
\begin{equation} 
\label{eq.sum_time_vis}
\rho^{vis}(\lambda,\beta,V_0,\omega_0, H_0) \\
   = \sum_j \rho_j^{vis}(\lambda,\beta,V_0,\omega_0, H_0),
\end{equation}
and their normalized sky plane residence time distribution density is:
\begin{equation}
\label{eq.normalized_residence_vis}
\rho_N^{vis}(\lambda,\beta,V_0,\omega_0, H_0) 
   = \frac{1}{C} \; \rho^{vis}(\lambda,\beta,V_0,\omega_0, H_0)
\end{equation}
where the normalization constant $C$ is from eq.~\ref{eq.constant}.  This normalization constant ensures that the normalized residence time in eq.~\ref{eq.normalized_residence_vis} is the fraction of all
possible particles, rather than the fraction of particles that satisfy
the constraints, so that we can determine the number density of
particles as derived below.

The
number density of particles with $H<H_0$, $V(H_0)<V_0$ and
$\omega<\omega_0$ as a function of sky plane location is (following the arguments in \S\ref{appendix.SkyPlaneNumberDensity})
\begin{eqnarray}
\nonumber
N^{vis}(\lambda,\beta,V_0,\omega_0, H_0) \;
  &=& \int^{H_0} \; dH \; n(\lambda,\beta,V_0,\omega_0,H)\Bigg|_{V(H_0)\; <\;    
      V_0, \ \omega\; <\; \omega_0} \\ 
  &=& \int^{H_0} \; dH \;  
      n_{HFD}(H) \; 
      {\rho_N^{vis}} (\lambda,\beta,V_0,\omega_0,H)\\
  \label{eq.num_vis}
\end{eqnarray}
\ie\ There are $N^{vis}(\lambda,\beta,V_0,\omega_0, H_0) \;
\Delta\lambda \; \Delta\beta$ particles in the range
$[\lambda-\Delta\lambda/2,\lambda+\Delta\lambda/2]$ and with
$[\beta-\Delta\beta/2,\beta+\Delta\beta/2]$ that also have
$H<H_0$, $V(H_0)<V_0$ and $\omega<\omega_0$.

The total number of particles in the sky that have $V(H_0) < V_0$,
$\omega < \omega_0$ and absolute magnitude brighter than $H_0$ is 
\begin{equation}
  = \int \dif\lambda
    \int \dif\beta \ 
      N^{vis}(\lambda,\beta,V_0,\omega_0,H_0)
\label{eq.num_vis_all}
\end{equation}
and the total number of particles in the range $\lambda_1 \le \lambda < \lambda_2$ and $\beta_1 \le \beta <
\beta_2$ with the same constraints on $V$, $\omega$ and
$H$ is
\begin{equation} \label{eq.time_range}
  = \int\limits^{\lambda_2}_{\lambda_1} \dif\lambda
    \int\limits^{\beta_2}_{\beta_1}     \dif\beta \ 
      N^{vis}(\lambda,\beta,V_0,\omega_0,H_0).
\end{equation}

\subsection{Infrared Detection of TCOs}
\label{appendix.InfraredDetection}

We convert between diameter ($D$) and
absolute magnitude using \citet{Fowler1992}:
\begin{equation}
\frac{D}{\mathrm{meters}} = \frac{1.329\times 10^6}{\sqrt{p_v}} 10^{-H/5}
\end{equation}
where $p_v$ is the albedo.  We used $p_v=0.11$ for the albedo
throughout this work to make the conversion from absolute magnitude
and diameter simple --- $H=18$ corresponds to 1\,km diameter and
factors of 10 changes in the diameter cause a change of 5 units in
absolute magnitude.

The observed IR flux ($F$) for a particle at geocentric distance $\Delta$
and heliocentric distance $r$ was calculated following
\citet{Harris2009} and \citet{Mainzer2011b}:
\begin{equation}
  F = \frac{\epsilon D^2}{4\Delta^2} 
        \int\limits_0^{2\pi} \dif\theta
        \int\limits_0^{\frac{\pi}{2}} \dif\phi
        \int\limits_{\lambda_1}^{\lambda_2}  \dif\lambda \;
          B(\lambda,T_{ss},\theta,\phi)\; \epsilon_b(\lambda)\; \sin\theta \; \cos\theta  
\label{eq.observedIRflux}
\end{equation}
where $\epsilon$ is a particle's thermal emissivity \citep[we used 0.9 per][]{Harris2009}, $\theta$ and $\phi$ are the longitude and
latitude on the object, $\epsilon_b$ is
the passband efficiency as a function of wavelength $\lambda$ (not
ecliptic longitude), $B$ is the blackbody flux for an object with
a sub-solar point temperature of $T_{ss}$.  That temperature is given by
\begin{equation}
T_{ss} = \left [\frac{S(r) \; (1-A)}{\eta  \epsilon \sigma} \right ]^{\frac{1}{4}}
\end{equation}
where $S(r)$ is the solar flux at heliocentric distance $r$, $\eta$ is
the beaming parameter \citep[we use $\eta=\pi$ for rapid rotators per][]{Harris2002}, $\sigma$ is the Stefan-Boltzmann constant, and $A$
is the bond albedo.

We assume that TCOs are in thermal equilibrium because they are 1) small, in
the range from $\sim$0.1\,m to $\sim$1.0\,m in diameter, 2) spend an average of about 9 months in Earth orbit with $r\sim1\au$, and 3) rotate relatively
quickly (see \S\ref{appendix.rotationRates}).  \ie\ there are no
longitudinal or latitudinal variations in the blackbody flux emitted
from the particle so that the function $B$ in eq.~\ref{eq.observedIRflux} can be removed from the integral.

The technique for calculating the total number of particles in the sky plane and their observable properties in the infrared is analogous to the techniques outlined above for optical surveys except that the flux, $F$, is used instead of apparent magnitude and we use the diameter distribution instead of the absolute magnitude distribution.

\subsection{Radar detection of TCOs}
\label{appendix.RadarSurveys}

The sky plane residence time distribution and implementations with constraints are readily extended to the
4-dimensions relevant to radar surveys for TCOs
in which a ground-based radar facility beams a radar signal and then searches for the reflected signal.  In this case the 4-dimensions are the ecliptic longitude and latitude as well as the
geocentric distance (range, $\Delta$) and geocentric speed
(range-rate, $\dot\Delta$).  

The ratio between the received ($P_{rec}$) and transmitted ($P_{tran}$) signal power at Earth is
\begin{equation}
  \frac{P_{rec}}{P_{tran}} = \frac{\epsilon_R \, A_R}{\Delta^4} 
\end{equation}
where $\epsilon_R$ is the
radar facility's detection efficiency and $A_R$ is the particle's
radar albedo (the fraction of the signal at the particle reflected
back to the transmitter).  We use a radar albedo of 0.15 typical for S-type asteroids and NEOs \footnote{\tt{http://echo.jpl.nasa.gov/$\scriptstyle\sim$lance/asteroid$\_$radar$\_$properties/nea.radaralbedo.html}}. The received power is detectable by the
facility if it exceeds the intrinsic detector and background noise.
However, the received power is spread over a range of wavelengths if
the particle is rotating, further degrading the signal.  The detected
signal-to-noise ratio is then \citep{Renzetti1988}:
\begin{equation}
 S/N = {{G_T \;G_A^2 \;\lambda^{\frac{5}{2}} \;A_R \;D\;\sqrt{NL}}
       \over
        {\Delta^4  \;k_b  \;T_N \;\sqrt{32 \;\omega}}}
\end{equation}
where $G_T$ is the peak transmitter gain, $G_A$ is the
antenna gain, $\lambda$ is the radar wavelength, $D$ is the target diameter, $N$ is the number of
observations, $k_b$ is
Boltzmann's constant, $T_N$ is the system noise temperature, $L$ is the
integration time, and $\omega$ is the object's rotation rate.

\subsection{Meteoroid rotation rates} 
\label{appendix.rotationRates}

Rotation rates of meteoroid-scale objects ($\sim$0.1 to 1\,m diameter)
are essentially unknown.  At the current time there is only one known
NEO with $H \ga 33$ (about 1\,m diameter) and the smallest object in the
Light Curve Database \citep[LCDB, ][]{Warner2009} has $H \sim 29.5$ (the
only known TCO, \RH).  The observational selection effects against
identifying the fastest and slowest rotators must be tremendous for
objects with $H \ga 33$ \ie\ the objects are so small that they are
faint and require relatively long exposure times so that it is
impossible to resolve fast rotation rates.  Observations of fireballs
add the complication of their interaction with the atmosphere and the
associated ablation effects that may modify a meteoroid's rotation
rate \citep{Beech2000}.  Periodic variations in the flux along a
meteor's trail on an image might represent the object's underlying
rotation period but these observations will specifically select those
objects with non-spherical shapes and fast rotation periods.  Thus,
there is not much that can be done to constrain the objects' rotation
rates except to use the available, limited and biased, observations.

\citet{Farinella1998} derived a spin period vs. diameter relationship of
$T=0.005\,{D\over\meter}$\,hours for kilometer-scale asteroids.  Extrapolating to
the meteoroid size range yields periods of 18\,s at 1\,m diameter and
2\,s at 10\,cm diameter.  While these rates seem fast,
\citet{Beech2000} point out that they are much slower than those
observed for meteors in that size range.  The meteor observations
suggest a period-diameter relationship of
$T\sim0.0001\,{D\over\meter}$\,hours  --- 50$\times$ faster than that suggested by \citet{Farinella1998}.  Their
results imply rotation periods of 0.5\,s at 1\,m diameter and
$0.05\second$ at $10\cm$ diameter.

\begin{figure}
\centering
\hspace*{0.712cm}
\ifgrayscale
\includegraphics[scale = 0.55]{rotationFrequency-vs-Diameter_gray-eps-converted-to.pdf}
\else
\includegraphics[scale = 0.55]{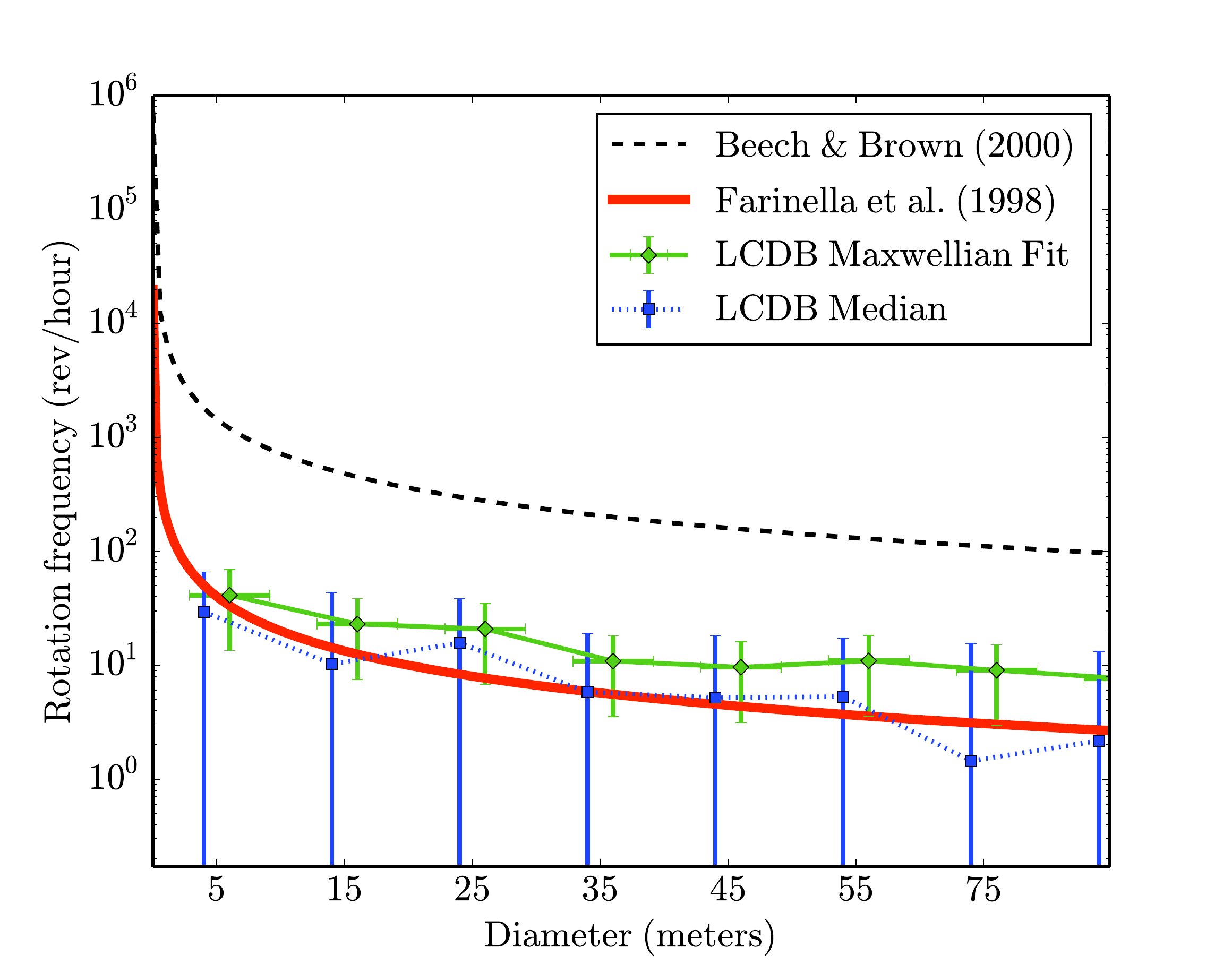}
\fi
\caption{Rotation frequency of small asteroids as a function of
  diameter from (thick solid line) \citet{Farinella1998} and (dashed
  line) \citet{Beech2000}.  The data points joined by dotted lines
  represent the median of the data in 10 meter diameter bins from the
  Light Curve Database \citep[LCDB;][]{Warner2009}.  The data points
  joined by thin solid lines represent our calculated median values
  from Maxwellian fits to the LCDB data in each bin.  The LCDB data
  was combined in $10\meter$ diameter bins and the data points are
  offset for the purpose of clarity by $\pm1\meter$ from the central
  values in each bin.}
\label{fig.rotationPeriod-vs-Diameter}
\end{figure}

The \citet{Farinella1998} period vs. diameter relationship agrees well with the median values from the LCDB in the range from $[0,100]$\,m diameter as illustrated in fig.~\ref{fig.rotationPeriod-vs-Diameter}.  We also fit the spin rates of the LCDB asteroids in the same intervals to Maxwellian distributions of the form:
\begin{equation}
\label{eq.rotationPeriodDistn}
f(\omega) 
  =  \sqrt{2\over\pi}\; {\omega^2\over\omega_0^3}\;
      \exp \Biggl( -\half \Biggl[ {\omega \over \omega_0} \Biggr]^2 \Biggr).
\end{equation}
The Maxwellian distribution is the expected form of the rotation
frequency distribution for a collisionally evolved asteroid population \citep[\eg][]{Pravec2002} but the median value from our fits ($\hat\omega \approx 1.54\;\omega_0$) is not a good representation of the predictions of \citet{Farinella1998} --- they are typically faster
than both the predicted and actual medians but approach the
predicted frequency as the diameter decreases.  Taken at face value it
implies that the smaller objects (near $10\meter$ diameter) have a more
Maxwellian spin rate distribution but the distribution becomes less
Maxwellian as the objects' diameters increase (to 100\,m diameter).
\citet{Pravec2008} have shown that the largest asteroids (many kilometers in diameter) have a Maxwellian spin-rate distribution but the spin-rates become less Maxwellian as their diameter decreases.  Since this work is concerned only with the smallest asteroids and, motivated by the results in fig.~\ref{fig.rotationPeriod-vs-Diameter}, we assume that 1)
the median rotation frequency ($\hat\omega$) of meteoroids in the $[0.1,10]$\,m diameter
size range is given by \citet{Farinella1998} and 2)
 their spin-rate
distribution is given by eq.~\ref{eq.rotationPeriodDistn} with $\omega_0=\hat\omega/1.54$.  

The smallest object for which a measured rotation period exists is the
TCO \RH\ with a rotation period of 165\,s \citep{Kwiatkowski2009}.  The median
rotation rate of objects of $2\meter$ diameter corresponding to the size of
\RH\ is about 1\,s and 36\,s using \citet{Beech2000} and
\citet{Farinella1998} respectively.  With these medians the
probability that a $2\meter$~diameter object has a rotation period of
$\ge165$\,s in the two models is $\sim$1\% and $\sim
2.2\times10^{-5}\,$\% respectively.  Thus, \RH's
rotation period would be a somewhat common $\sim2.6-\sigma$ event using our technique but a $~5.2-\sigma$ event if the asteroids in that size range had a
median given by \citet{Beech2000}.


\bibliographystyle{icarus}
\bibliography{./references}


\clearpage
\begin{table}[htdp]
\begin{center}
\begin{tabular*}{\textwidth}{lcrrrrllr}
\hline
Survey & Type & FOR       & $T_{exp}$ & $V_{lim}$ & $\omega_{lim}$ & $N_{sky}$ & $N_{TCO}$ & $\tau_{ref}$ \\
       &      & (deg$^2$) &  ($\second$)    &           & $\arcdeg/\days$        &           & TCO/$\days$   & $\days$             \\
\hline
\hline
PS1\tnm{1}        & wide-area &   1,000 & 40 & 21.7 &  3  & 0.03    & 0.013 &   2.3  \\
ATLAS\tnm{2}      & all-sky   &  20,000 & 30 & 20.0 & 15  & 0.0084  & 0.01  &   0.8  \\
SST\tnm{3}        & all-sky   &  20,000 &  5 & 22.0 & 20  & 0.4     & 0.02  &  18.0 \\
LSST\tnm{4}       & wide-area &   7,000 & 15 & 24.7 & 10  & 8.5     & 0.27  &  31.0  \\
\hline
Subaru-HSC\tnm{5} & targeted  &   450   & 15 & 24.5 & 1.0 & 0.4$^{\dagger}$     & 0.05  &   8.0  \\
\hline
CAMO\tnm{6}       & meteor    & 20,000 & N/A & 7.5 & N/A   &   $\sim0$  & N/A  &  N/A  \\
CAMS\tnm{7}       & meteor    & 20,000 & 0.017 & 4.8 & 11,300   &   $\sim0$  & 0.04  &  26.1  \\
\hline
\end{tabular*}
\end{center}
\caption{TCO detection performance for seven optical surveys.  For the
  first five telescopic surveys we consider only TCOs with absolute
  magnitude $H<38$ corresponding roughly to those $>10\cm$ diameter.
  FOR is the Field-of-Regard --- the total average amount of sky
  surveyed each night in a mode suitable for identifying TCOs.
  $V_{lim}$ and $\omega_{lim}$ are the survey's limiting $V$-band
  magnitude and approximate rate of motion.  $N_{sky}$ and $N_{TCO}$
  are the number of TCOs on the sky that are detectable by the survey
  at any instant and the average number of TCOs detectable by the
  survey per day.  $\tau_{ref}$ is the refresh rate for the observable
  TCOs, equivalent to their average observable lifetime.}
\tnt{1}{\citet{Denneau2013}} 
\tnt{2}{\citet{Tonry2011}}
\tnt{3}{\citet{Monet2013}} 
\tnt{4}{\citet{Ivezic2008}}
\tnt{5}{\citet{Takada2010}} 
\tnt{6}{\citet{Weryk2008}}
\tnt{7}{\citet{Jenniskens2011}} 
\tnt{\dagger}{In this case only, the number of TCOs detectable 
              by the HSC survey in one night.}
\label{t.TCOSurveyPerformance}
\end{table}


\clearpage

\clearpage

\clearpage

\clearpage

\clearpage

\clearpage

\clearpage

\clearpage

\clearpage

\clearpage

\clearpage

\clearpage

\clearpage

\clearpage

\end{document}